\documentclass[12pt]{article}

\usepackage{epsf}

\usepackage{graphicx}%
\usepackage{color}

\def\refitem#1{\relax}

 \newfont{\cyrfnt}{wncyr9 scaled 1120}

\begin{document}

\begin{center}
{\bfseries Hadron Resonance Gas Model for An Arbitrarily Large Number of  Different Hard-Core Radii}

\vskip 5mm

K. A. Bugaev$^{1 *}$, V. V. Sagun$^{1, 2}$,  A. I. Ivanytskyi$^{1}$, I. P. Yakimenko$^{3}$, 
E. G. Nikonov$^{4}$,  A.V. Taranenko$^{5}$
and G. M. Zinovjev$^{1}$

\vskip 5mm

{\small
$^1${\it
Bogolyubov Institute for Theoretical Physics, Metrologichna str. 14$^B$, Kiev 03680, Ukraine
}
\\
$^2${\it
CENTRA, Instituto Superior T$\acute{e}$cnico, Universidade de Lisboa,
Av. Rovisco Pais 1, 1049-001 Lisboa, Portugal
}
\\
$^{3}${\it
Department of Physics, Chemistry and Biology (IFM), Link\"oping University, SE-58183 Link\"oping, Sweden
}
\\
$^{5}${\it
Laboratory for Information Technologies, JINR, Joliot-Curie str. 6, 141980 Dubna, Russia
}
\\
$^5${\it  National Research Nuclear University ``MEPhI'' (Moscow Engineering Physics
Institute), Kashirskoe Shosse 31, 115409 Moscow, Russia}
\\
$*$ {\it
E-mail: Bugaev@th.physik.uni-frankfurt.de
}
}
\end{center}

\vskip5mm

\centerline{\bf Abstract}
{ \small 
We develop a novel formulation of the hadron-resonance gas model which, besides a hard-core repulsion, explicitly 
accounts for the surface tension induced by the interaction between the particles. Such an equation of state allows us to 
go beyond the Van der Waals approximation for any number of different hard-core radii. 
A comparison with the Carnahan-Starling  equation of state shows that the new model is valid for packing fractions 
0.2-0.22, while the usual Van der Waals model
is inapplicable at packing fractions above 0.1-0.11.  Moreover, it is shown that the equation of state with 
induced surface tension is softer than the one of hard spheres and remains causal at higher particle densities. 
The great advantage of our model is that there are only two equations to be solved and it does not depend on 
the various values of the hard-core radii used for different hadronic resonances.  
Using this novel equation of state we obtain a high-quality fit of the ALICE
hadron multiplicities measured at center-of-mass energies of 2.76 TeV per nucleon.
Furthermore, using the traditional hadron-resonance gas model with multi-component hard-core repulsion 
and the novel one 
we investigate  the  recently suggested  model
in which the proper volume of a hadron is proportional to its mass.  
We find that the high-temperature maximum of $\chi^2/ndf$ observed in the latter model  always 
appears in the region located far above  the limit of its applicability.

\vskip5mm

\section{Introduction} 

During the last few years the hadron-resonance gas model (HRGM) \cite{PBM06} became a precise and 
reliable tool to extract the parameters of chemical freeze-out (CFO). 
The traditional versions of the HRGM are basically the Van der Waals equation of state (EOS) which is 
formulated for all hadrons and hadronic resonances and which employs either  one hard-core radius $R$ for all hadrons, 
or at least  two hard-core radii: one for mesons $R_m$ and one for baryons $R_B$. 
However, the main achievements  in this field come from a formulation of the HRGM with multi-component  hard-core 
repulsion  (MHRGM) \cite{Horn,SFO,Bugaev:2014,Veta14}.  At the moment,  the MHRGM contains at most five 
different hard-core radii:
the pion hard-core radius $R_\pi$, the kaon hard-core radius  $R_K$, the hard-core radius of 
$\Lambda$-(anti)hyperons in addition to the ones for other baryons $R_B$ and for other mesons $R_m$. 
Having introduced four hard-core radii (no radius for $\Lambda$-(anti)hyperons) it was possible for the first time 
to describe the Strangeness Horn \cite{Horn} with the highest quality $\chi^2/ndf \simeq 7.3/14$ and at the same time 
successfully fit 111 independent hadron multiplicities ratios measured  from  AGS ($\sqrt{s_{NN}}=2.7$ GeV) to 
top RHIC   ($\sqrt{s_{NN}} = 200$ GeV) collision  energies with  the fit quality  $\chi^2/ndf \simeq 1.16$, which was 
the best result at that time.  

A couple of years later, taking into account one more global fitting parameter -- namely  the hard-core radius of 
$\Lambda$-(anti)hyperons, which turned out to be a very influential parameter -- 
it was possible to  decrease the value of  $\chi^2/ndf $ by about 20 percent and to reach the best fit quality of these 
data which is   $\chi^2/ndf \simeq 0.95$ \cite{Veta14}.  Furthermore, the developed  MHRGM  allowed 
one to formulate the concept of separate CFOs of strange and non-strange hadrons \cite{SFO,BugaevIndia}, 
which naturally explains the apparent chemical non-equilibrium of strange charge in nucleus-nucleus collisions.  
And last but not least, 
the very high quality description of data allowed us to study the thermodynamics at CFO  with very high confidence and 
to find novel irregularities and signals of mixed-phase formation in nuclear collisions \cite{Bugaev:2014,Bugaev:2015}. 

In view of  future experiments at the NICA-JINR and FAIR-GSI accelerators we expect to obtain much more 
experimental data with a substantially  higher accuracy. Evidently, these new data will, hopefully, allow us to study 
the second virial coefficients of the most abundant (or maybe even those of all measured) hadrons. For such a task a 
MHRGM with many hard-core radii should be formulated. However, since 
the existing MHRGM for $N$ different hard-core radii  requires the knowledge of  a solution of $N$ transcendental 
equations, a further 
increase of  the number of hard-core radii (i.e., $N\sim 100$, corresponding to the various hadronic species created
in a collision) will lead to a substantial increase of computational time and will destroy 
the main attractive feature of the thermal model, i.e., its simplicity. 

Recently, there appeared a work \cite{Vovch15} in which the authors suggested to use the relation 
$V_h = C \cdot m_h\,$ between the hadronic proper volume $V_h$  and its mass $m_h$. However, 
the authors of Ref.  \cite{Vovch15}  applied the 
Van der Waals approximation, which rapidly becomes invalid when the hard-core radii become too large.
Since the issues raised in Ref.\ \cite{Vovch15} are very important for heavy-ion phenomenology, we would like to 
re-analyse ALICE data  
\cite{Abelev:2013vea,Abelev:2013zaa,Abelev:2013xaa,Knospe:2013tda,Adam:2015vda,Donigus:2015bsa,Adam:2015yta} 
with the conventional MHRGM and with the novel model, which is able to go beyond the usual Van der Waals  
approximation.

For this purpose, we present here an entirely new version of the HRGM with multi-component hard-core repulsion: 
The Van der Waals EOS with induced surface tension (abbreviated IST EOS in the following).  The IST EOS is based on 
the virial expansion for a multi-component mixture and, hence,  it naturally switches between the low- and high-density limit. 
Comparing it to the Carnahan-Starling EOS \cite{CSeos} for one and two particle species we find an almost perfect 
agreement between them up to packing fractions $\eta \simeq 0.2-0.22$. 
Its great advantage is that, independent from the number of  different hard-core radii, the IST EOS only involves
solving a system of two transcendental equations.  
Using the IST EOS we successfully  fit the ALICE data \cite{Abelev:2013vea,Abelev:2013zaa,Abelev:2013xaa,Knospe:2013tda,Adam:2015vda,Donigus:2015bsa,Adam:2015yta} and compare the obtained results with the results found  
by the  MHRGM as well as by the model of Ref.\ \cite{Vovch15}. 

This paper is organized as follows: in the next section we present the main features of the  IST EOS;
in Sec.\ 3 the fit of the ALICE data by different versions of the HRGM is described in detail; 
Sec.\ 4 is devoted to a thorough  analysis of the applicability bounds of various versions 
of the HRGM, including 
 the Vovchenko-Stoecker model  \cite{Vovch15}; the discussion of results is given in Sec.\ 5; 
our conclusions are formulated in Sec.\ 6, and the Appendix contains useful formulae. 

\section{HRGM with the induced surface tension}

In order to use an arbitrarily large number of independent hard-core radii, we 
 employ  the IST EOS which is much  more effective compared to  the traditional  MHRGM  
 \cite{Horn,SFO,Bugaev:2014,Veta14}.
Such a model was derived on the basis of the virial expansion for a  multi-component mixture \cite{Bugaev:13NPA} 
obtained for the simplified  statistical multifragmentation model  \cite{Mekjian} with an infinite number of 
hard-core radii of nuclear fragments of all sizes. 
The thermodynamically consistent  equation of state developed in Ref.\ \cite{Bugaev:13NPA} is a system of coupled  
equations for the pressure  $p$ and  the induced surface-tension coefficient $\Sigma$. 
Applying  it to  the pressure and  the induced surface-tension coefficient of the hadron-resonance gas, we obtain
\begin{eqnarray}
\label{EqI}
p &=& T \sum_{k=1}^N \phi_k \exp \left[ \frac{\mu_k}{T} - \frac{4}{3}\pi R_k^3 \frac{p}{T} - 4\pi R_k^2 \frac{\Sigma}{T} \right]
\,, \\
\label{EqII}
\Sigma &=& T \sum_{k=1}^N R_k \phi_k \exp \left[ \frac{\mu_k}{T} - \frac{4}{3}\pi R_k^3 \frac{p}{T} - 4\pi R_k^2 \alpha 
\frac{\Sigma}{T} \right] \,,\\
\label{EqIII}
\mu_k &=& \mu_B B_k + \mu_{I3} I_{3k} + \mu_S S_k \,,
\end{eqnarray}
where $\mu_B$, $\mu_S$, $\mu_{I3}$ are the baryonic,  the strange, and the third projection of  the 
isospin  chemical potential,  respectively. Here
$B_k$, $S_k$, $I_{3k}$, $m_k$ and $R_k$ denote, respectively,  the corresponding charges,  mass, 
and hard-core radius of the $k$-th hadronic species. The sums in  Eqs.\ (\ref{EqI}) and (\ref{EqII})  run
over all hadronic species; their corresponding antiparticles are 
considered as independent species.

The one-particle thermal density $\phi_k$ in Eqs.\ (\ref{EqI}) and (\ref{EqII})  accounts for  the  Breit-Wigner  
mass attenuation and is written in the Boltzmann approximation
\begin{eqnarray}
\label{EqIV}
\phi_k = g_k  \gamma_S^{|s_k|} \int\limits_{M_k^{Th}}^\infty  \,  \frac{ d m}{N_k (M_k^{Th})} 
\frac{1}{(m-m_{k})^{2}+\Gamma^{2}_{k}/4} 
\int \frac{d^3 p}{ (2 \pi)^3 }   \exp \left[ -\frac{ \sqrt{p^2 + m^2} }{T} \right] \,,
\end{eqnarray}
where $g_k$ is the  degeneracy factor of the $k$-th hadronic species,
$\gamma_S$ is the strangeness suppression factor \cite{Rafelski}, $|s_k|$ is the number of valence  
strange quarks and antiquarks in this hadron species,
${N_k (M_k^{Th})} \equiv \int\limits_{M_k^{Th}}^\infty \frac{d m}{(m-m_{k})^{2}+\Gamma^{2}_{k}/4} $ denotes 
a normalization factor, while $M_k^{Th}$ corresponds to the decay threshold mass of the $k$-th hadronic species.

To study nuclear collisions the system of Eqs.\  (\ref{EqI}), (\ref{EqII}), and  (\ref{EqIII}) should be supplemented by 
the strange-charge conservation law
\begin{eqnarray}
\label{EqV}
  \phi_k S_k \exp \left[ \frac{\mu_k}{T} - \frac{4}{3}\pi R_k^3 \frac{p}{T} - 4\pi R_k^2 \frac{\Sigma}{T} \right] = 0 \,,
\end{eqnarray}
which completes the  system of equations of the IST EOS.

The dimensionless  parameter $\alpha>1$ is  introduced  in Eq.\ (\ref{EqII}) due to the freedom of the Van der Waals 
extrapolation to high densities \cite{Bugaev:13NPA}.  As it is shown below,  the  parameter $\alpha$ 
makes the  Van der Waals  EOS  more realistic in the high-density limit.
In principle, $\alpha$ can be a regular function of $T$ and $\mu$,  however, for the sake of simplicity it is fixed to 
a constant value. 
In the work  \cite{Bugaev:13NPA} it was established that the parameter  $\alpha$  should obey the inequality  
$ \alpha >1 $ in order to reproduce the physically correct phase-diagram properties of nuclear matter.  

The physical meaning of  $\alpha$ can  be revealed,  if we  use the following relation
\begin{eqnarray}
\label{EqVI}
\Sigma_k & =   &  p_k R_k  \, \exp \left[ - 4 \pi R_k^2 \cdot (\alpha-1)\, \frac{\Sigma}{T} \right] \, ,
\end{eqnarray}
between the partial pressure $p_k$ of the $k$-th hadronic species
and the corresponding partial  induced surface-tension coefficient $\Sigma_k$. The system pressure  
$p = \sum_{k=1}^N p_k$ and the total induced surface-d tension coefficient $\Sigma = \sum_{k=1}^N \Sigma_k$ are 
the sums of their corresponding partial values.

In the next paragraph, for the sake of simplicity,  we assume that all particles have the same hard-core radius 
$R_k = \tilde R$ for $ k = 1, 2, 3, \ldots$ and, consequently, the same proper volume $\tilde v$.
Using Eq.\ (\ref{EqVI}) one can identically  rewrite  Eq.\ (\ref{EqI}) in the form
\begin{eqnarray}
\label{EqVII}
p & =   & T \sum_{k=1}^N \phi_k
\exp\left[ \frac{{\mu_k}}{T} - \tilde v \frac{p}{T}- 3 \tilde v  \frac{p }{T} \cdot \exp \left[ -3 \tilde v \cdot (\alpha-1)\, 
\frac{\Sigma}{T\, \tilde R} \right] \right] \nonumber \\
& =   & T \sum_{k=1}^N \phi_k  \exp\left[ \frac{{\mu_k}}{T} - v^{eff}  \frac{p}{T}\right]   \,,
\end{eqnarray}
where we introduce  the effective excluded volume of hadrons of species $k$
\begin{eqnarray}
\label{EqVIII}
v^{eff}  & =   & \tilde  v \left[1 +  3  \cdot \exp \left[ -3 \tilde v  \cdot (\alpha-1)\, \frac{\Sigma}{T\, \tilde R} \right]\right] \,.
\end{eqnarray}
The low-density limit is obtained for $\mu_k \rightarrow - \infty$. In this limit  
$\frac{\Sigma \tilde v}{T \tilde R} \rightarrow 0$  and, 
hence,  $v^{eff}  \simeq 4 \tilde  v $, i.e., it correctly reproduces the excluded volume for the one-component case.  
In the  high-density limit 
$\frac{\Sigma \tilde v}{T \tilde R} \gg 1$,  since $\mu_k/T \gg 1$. Therefore, for $\alpha >1$ the exponential function 
on the right-hand side of  Eq.\ (\ref{EqVIII}) vanishes and  
the effective excluded volume becomes equal to the proper volume, i.e., $v^{eff} \simeq  \tilde v $. Thus, in the 
present model the parameter  $\alpha$ switches  between the excluded-volume and the proper-volume regimes.
In order to apply this equation of state to describe the hadron multiplicities we have  to fix the value of  $\alpha$.
For this purpose we  compare the one-component system of equations of state (\ref{EqI}), (\ref{EqII}) with the 
famous one-component Carnahan-Starling (CS) EOS \cite{CSeos},
\begin{eqnarray}
\label{EqXV}
P~=~\rho\, T  \, Z_{CS}, 
\hspace*{0.2cm} Z_{CS} = \frac{1+\eta +\eta^2 - \eta^3}{(1 - \eta)^{3}} \,.
\end{eqnarray}
The CS EOS 
is more accurate  at higher densities than the  excluded-volume model  (EVM),  which is a more proper  
name for  the Van der Waals EOS with repulsion. 
Here the packing fraction of particles of the same hard-core radius $R$  is $\eta =  \frac{4}{3} \pi R^3 \, \rho$, 
where $\rho$ is   their  particle density. From  Fig.\  \ref{Fig1}
one can see that up to  
$\eta \simeq$ 0.22 the IST EOS  with $\alpha=$1.25 reproduces both the compressibility factor $Z$ and the 
speed of sound $c_S$ of the CS EOS.  Figure  \ref{Fig1} represents an extreme case, when the pion hard-core radius is 
set to zero, while the baryons (here, nucleons and $\Delta_{P33}(1232)$ isobars were accounted for) have a rather 
large radius $R_B = 0.4$ fm, and additionally it corresponds to an unphysically high temperature $T=200$ MeV.  
To include point-like pions, we added the ideal pion-gas pressure to the baryonic pressure given by Eq.\ (\ref{EqXV}) 
according to Ref.\ \cite{SSpeed}.
Therefore,  the value $\alpha =1.25$ is used in the present work. 

We have to note that for pions we used the Bose-Einstein distribution function 
\begin{eqnarray}
\label{EqX}
\phi_\pi = g_\pi  \int \frac{d^3 p}{ (2 \pi)^3 }   \frac{1}{\exp \left[ \frac{ \sqrt{p^2 + m_\pi^2} }{T} \right]  - 1} \,,
\end{eqnarray}
instead of Eq.\ (\ref{EqIV}) because at high temperatures the quantum correction cannot be ignored. 
In principle, one would need the quantum generalization of the system  (\ref{EqI}), (\ref{EqII}), but the fit of 
ALICE data corresponds
to vanishing values of all chemical potentials. In addition, the pion hard-core radius is rather small compared to 
other mesons, therefore, for the sake of simplicity,  in Eqs.\ (\ref{EqI}), (\ref{EqII}), and  (\ref{EqIII}) we used  
Eq.\ (\ref{EqX}) instead of the Boltzmann approximation  (\ref{EqIV}).

It is worth noting that the traditional EVM, i.e.,  the Van der Waals  EOS,   corresponds to the case 
$\alpha = 1$. From Fig.\ \ref{Fig1} one can see that such an EOS is in agreement with the CS EOS 
up to a packing fraction $\eta \simeq 0.1-0.11$  only.  
Also from this  figure it is clearly seen that the EVM violates causality at a baryonic density of about 
$0.64$ fm$^{-3}$ ($\eta \simeq 0.17$), while the IST EOS remains causal for baryonic densities above 1 fm$^{-3}$. 
Such an advantage of the IST EOS is of principal  importance  for hydrodynamic simulations.

\begin{figure}[htbp]
\centerline{\hspace*{-10mm}
\includegraphics[width=104mm]{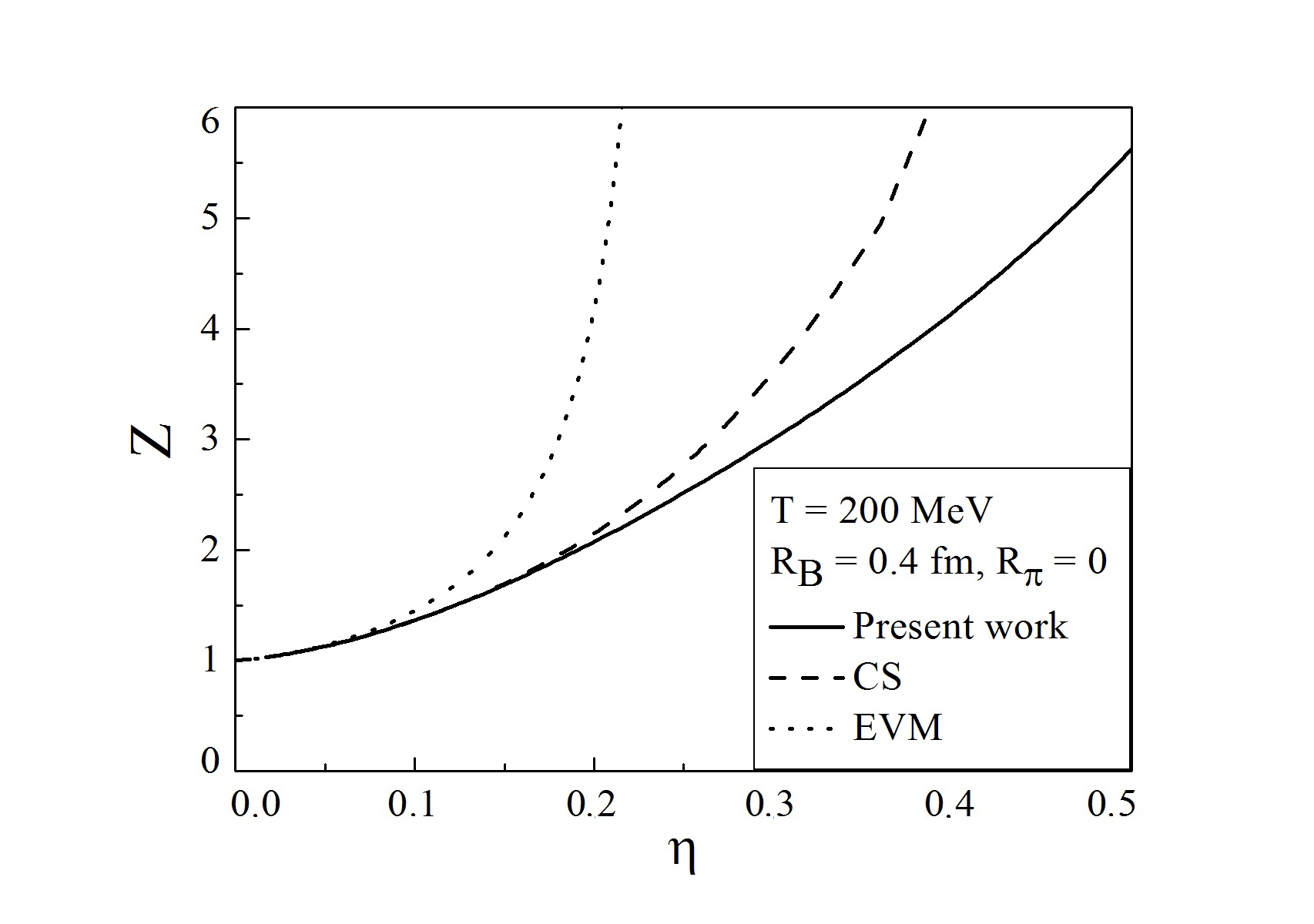}
  \hspace*{-10mm}
\includegraphics[width=91mm]{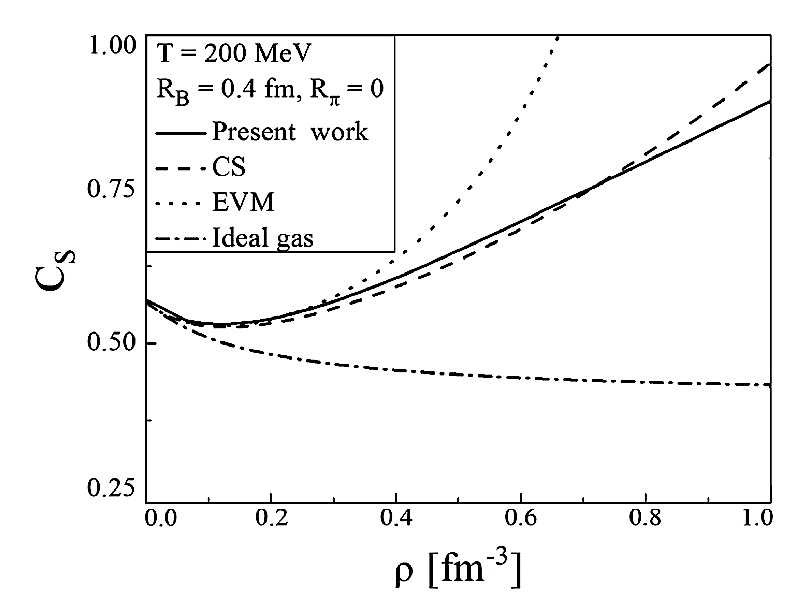}
 } 
 \caption{{\bf Left panel:} The compressibility factor $Z$  of  a gas consisting of point-like pions and baryons 
 (nucleons and $\Delta_{P33}(1232)$ isobars were accounted for) with a hard-core radius of $0.4$ fm  is shown for 
 different EOS  as a function of baryon packing fraction $\eta$.  The  Van der Waals  EOS (dotted curve), the 
 IST EOS (solid curve),  and the CS EOS (long dashed curve) are shown 
  for $T=$ 200 MeV. {\bf Right panel:} The speed of sound as a function of baryonic density is shown for the 
  same EOS as in the left panel  and with the same notations. The dotted-dashed curves shows the  
   speed of sound of point-like pions and baryons.}
  \label{Fig1}
\end{figure}

Using  the value $\alpha =1.25$  we employed   the IST EOS to fit  the  independent  hadronic-multiplicity ratios 
\cite{VetaNEW} measured  in central nuclear collisions  for center-of-mass energies 
$\sqrt{s_{NN}} = $ 2.7, 3.3, 3.8, 4.3, 4.9, 6.3, 7.6, 8.8, 9.2, 12, 17, 62.4, 130, and 200 GeV (for details see 
Ref.\ \cite{Horn,SFO,Veta14}).
Then we compared the obtained results with  the ones  from the  MHRGM with hard-core radii found   in 
Ref.\ \cite{Veta14}, i.e., for
the  hard-core radii of baryons  $R_{b}$=0.355 fm, mesons $R_{m}$=0.4 fm, pions $R_{\pi}$=0.1 fm, kaons
$R_{K}$=0.395 fm, and $\Lambda$ hyperons $R_{\Lambda}$=0.11 fm. 
Treating $\gamma_S$ as a fitting  parameter  in the IST EOS  
we found that the best data description of the same  data set  which was used  in Ref.\ \cite{Veta14} corresponds
to the following  values of 
hard-core radii (new radii hereafter)  of baryons $R_{b}$=0.365 fm, mesons $R_{m}$=0.42 fm, 
pions $R_{\pi}$=0.15 fm, kaons $R_{K}$=0.395 fm, and $\Lambda$ hyperons $R_{\Lambda}$=0.085 fm. 
These values of the hard-core radii generate $\chi^2/ndf=57.099/55 \simeq 1.038$ which is about 9\% larger than 
the $\chi^2/ndf$ found earlier in Ref.\ \cite{Veta14} for the EVM. 
Hence, in the present work we use exactly this new  set of hard-core  radii for the analysis of the ALICE data. 

Compared to the values found by the MHRGM, one sees that only the pion hard-core radius increased by 50\%, 
while the hard-core radius of  $\Lambda$ hyperons diminished by 20\%.  The most important thing is that  these  
radii remain essentially smaller than $R_{b}$, $R_{m}$, and $R_{K}$.  The latter hard-core radii are practically unchanged.
The most prominent changes of the fit are compared  in Fig.\ \ref{Fig2} with the corresponding values from 
Ref.\ \cite{Veta14}. 
As one can see from this figure, at the center-of-mass energy $\sqrt{s_{NN}} = 6.3$ GeV only the ratio $K^-/K^+$ is 
essentially improved, while 
for $\sqrt{s_{NN}} = 130$ GeV   the ratios $K^+/\pi^+$,  $\bar p/\pi^-$,  $\Omega/\pi^-$, and $\phi / K^-$ are described 
better than within the MHRGM. 
The  fit results  obtained by the  MHRGM and  by the IST EOS for the other collision energies 
are hardly distinguishable from each other.

\begin{figure}[htbp]
\centerline{
\includegraphics[width=77mm]{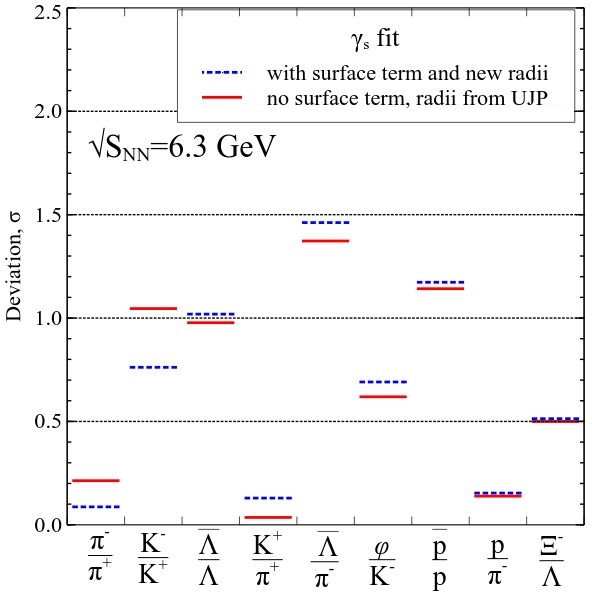}
  \hspace*{0.22cm}
\includegraphics[width=77mm]{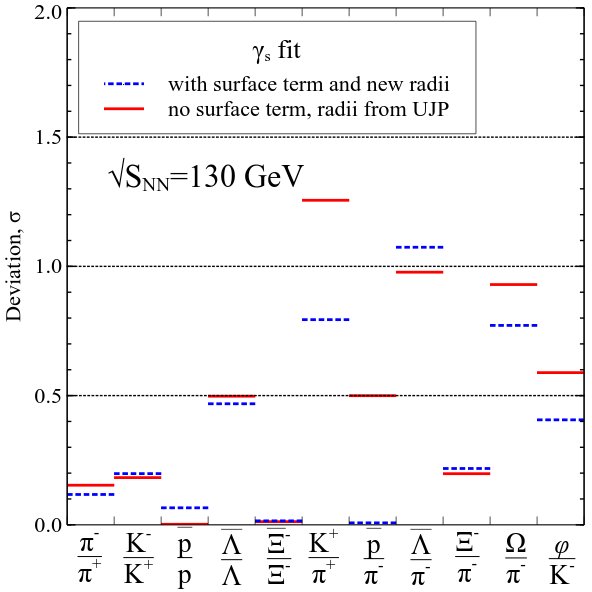}
  }
 \caption{(Color online) Deviation of theoretically predicted hadronic ratios from experimental values in units of
 experimental error $\sigma$ for $\sqrt{s_{NN}} = 6.3$ GeV (left panel) and  $\sqrt{s_{NN}} = 130$ GeV 
(right panel). Solid lines correspond to the original MHRGM with $\gamma_{S}$ fit \cite{Veta14}, while 
dashed lines correspond to the IST EOS  fit with the $\gamma_{S}$ parameter \cite{VetaNEW}.
 }
\label{Fig2}
\end{figure}

\section{Hadron-Multiplicity Data of the ALICE Experiment}

The data to be fitted are the hadron multiplicities at midrapidity $\frac{dN}{dy}|_{|y| < 0.5}$, measured by the 
ALICE detector at $\sqrt{s_{NN}} = 2.76$ TeV in Pb + Pb collisions. All of them are $p_T$-integrated. 
The detector cannot measure particles at low $p_T$; as a result, the multiplicities have to be extrapolated. 
The extrapolation was  performed by a blast-wave fit in each centrality bin; the extrapolation error is included in the 
systematic errors. Thus,  all the following data are  from the same kinematic region.

In general we are  interested in the most central collisions. However, the experimental centrality bins are different for 
different particles. If there is enough statistics there are more bins; whereas for the rare particles there are less bins 
and the bins are larger. In some cases the hadron multiplicity does not depend on centrality, but for most hadron species 
one has to be very careful with the centrality selection. Simultaneous fitting of, e.g.\ the $\pi$ multiplicity at 0-5\% centrality, 
$\Xi$ at 0-10\% centrality, and $^{3}He$ at 0-20\% centrality, is not acceptable  if one aims at a precise description of the
data. On the other hand, the multiplicity ratios (with corresponding centrality bins) seem to be independent of centrality. 
Therefore, first we give a table of raw experimental data from publications (see Table \ref{tab:exp_data}) and then a table of 
quantities  which we used for a subsequent analysis  (Table \ref{tab:rat_fit}). 
Our experience tells us that, 
besides the numerical convenience,  fitting the ratios of hadron yields  provides a better stability of the results.

Similarly to Ref.\  \cite{Stachel:2013zma} we 
do not include the $K^*$ data into the fitting procedure. The $K^*$ yield is known to have around 
$3\sigma$ deviation from the thermal-model prediction, and it is not included to fit the thermal parameters in 
recent works (see caption of Fig.\ 1 in Ref.\ \cite{Stachel:2013zma}), because ``as a strongly decaying resonance its 
yield can be significantly modified after chemical freeze-out'' \cite{Stachel:2013zma}. We  agree with this argument,  
since the reactions like $K+\pi \leftrightarrow K^*$ can occur after chemical freeze-out and change the $K^*$ yields.
Due to a technical reason  we do not fit the $K_S^0$ data. As one can see from Table 1 the multiplicity of  $K_S^0$ meson is identical to the ones of $K^+$ and $K^-$ mesons. Since in our fit 
 all chemical potentials are fixed to zero (see below), then the $K_S^0$ meson  is
formally indistinguishable from  $K^+$ and $K^-$ mesons and because of this fact any ratio involving  $K_S^0$ meson is not
independent. Therefore, it is omitted from the fit.

Hence, overall we have  19 hadron species to fit ($\pi^{\pm}$, $K^{\pm}$, $p$, $\bar{p}$, $\phi$,  $\Lambda$, 
$\Xi^{\pm}$, $\Omega^{\pm}$, $d$, $\bar{d}$, $^3He$, $^3\overline{He}$, $^4\overline{He}$, $^{3}_{\Lambda}H$, 
$^{3}_{\bar{\Lambda}} \bar{H}$), or 18 ratios,  respectively.  
In Ref.\ \cite{Stachel:2013zma} the centrality is 0-10\% for 
all species, but in  Refs.\
\cite{Abelev:2013vea,Abelev:2013zaa,Abelev:2013xaa,Knospe:2013tda,Adam:2015vda,Donigus:2015bsa,Adam:2015yta}
such a centrality class is not always present, as one can see from Table  \ref{tab:exp_data}. Fortunately, the multiplicity 
ratios are rather independent on centrality -- if the same centrality class is taken for numerator and denominator. 
Consequently, we take such ratios for our fit. To obtain the 0-10\% centrality class from the 0-5\% and 5-10\% classes 
we take the average value of the corresponding multiplicities and the average values of the errors. Apparently, this 
can be done under assumption that the relative experimental error of $dN/dy$ is independent of  centrality, 
which is approximately true for the ALICE data.
Therefore,  constructing the ratio $r = \frac{A}{B}$ from species $A + \Delta A$ and $B + \Delta B$, we estimate a 
relative error $\epsilon_r \equiv \frac{\Delta r}{r}$ as $\sqrt{\epsilon_A^2 + \epsilon_B^2}$ \cite{Taylor82}, where 
$\epsilon_{A} = \frac{\Delta A}{A}$ and $\epsilon_{B} = \frac{\Delta B}{B}$. This is certainly overestimating the error,
because a part of the systematic errors, such as errors related to detector acceptance, is usually canceled  in the 
experimentally measured ratios. If statistical and systematic errors $\Delta A_{stat}$ and $\Delta A_{sys}$ are given, 
we add them as $\Delta A = \sqrt{\Delta A_{stat}^2 + \Delta A_{sys}^2}$ \cite{Taylor82}.

\begin{table}
\begin{center}
  \begin{tabular}{cccc}
                  &  dN/dy or ratio                            & centrality            &        Ref.                   \\
    \hline
    $\pi^+$       &  $733 \pm 54$, $606 \pm 42$, $ 455 \pm 31  $   & 0-5, 5-10, 10-20\%    &  \cite{Abelev:2013vea}    \\
    $\pi^-$       &  $732 \pm 52$, $604 \pm 42$, $ 453 \pm 31  $   & 0-5, 5-10, 10-20\%    &  \cite{Abelev:2013vea}    \\
    $K^+$         &  $109 \pm  9$, $ 91 \pm  7$, $  68 \pm  5  $   & 0-5, 5-10, 10-20\%    &  \cite{Abelev:2013vea}    \\
    $K^-$         &  $109 \pm  9$, $ 90 \pm  8$, $  68 \pm  6  $   & 0-5, 5-10, 10-20\%    &  \cite{Abelev:2013vea}    \\
    $p$           &  $ 34 \pm  3$, $ 28 \pm  2$, $  21 \pm  1.7$   & 0-5, 5-10, 10-20\%    &  \cite{Abelev:2013vea}    \\
    $\bar{p}$     &  $ 33 \pm  3$, $ 28 \pm  2$, $21.1 \pm  1.8$   & 0-5, 5-10, 10-20\%    &  \cite{Abelev:2013vea}    \\
    $\Xi^-$       &  $ 3.34 \pm 0.06 \pm 0.24$, $2.53 \pm 0.04 \pm 0.18$ & 0-10, 10-20\%   &  \cite{Abelev:2013zaa}    \\
    $\Xi^+$       &  $ 3.28 \pm 0.06 \pm 0.23$, $2.51 \pm 0.05 \pm 0.18$ & 0-10, 10-20\%   &  \cite{Abelev:2013zaa}    \\
    $\Omega^-$    &  $ 0.58 \pm 0.04 \pm 0.09$, $0.37 \pm 0.03 \pm 0.06$ & 0-10, 10-20\%   &  \cite{Abelev:2013zaa}    \\
    $\Omega^+$    &  $ 0.60 \pm 0.05 \pm 0.09$, $0.40 \pm 0.03 \pm 0.06$ & 0-10, 10-20\%   &  \cite{Abelev:2013zaa}    \\
    $\Lambda$     &  $ 26 \pm  3$, $22 \pm 2$                      & 0-5, 5-10\%           &  \cite{Abelev:2013xaa}    \\
    $K_S^0$       &  $110 \pm 10$, $90 \pm 6$                      & 0-5, 5-10\%           &  \cite{Abelev:2013xaa}    \\
    $\frac{\phi}{K^-} \times 4$ &  $0.45 \pm 0.1$                  & 0-10\%                &  \cite{Knospe:2013tda}    \\
    $K^*(892)/K^-$              &  $0.2  \pm 0.05$                 & 0-10\%                &  \cite{Knospe:2013tda}    \\
    $d$           & $(9.82 \pm 0.04 \pm 1.58)\cdot 10^{-2}$        & 0-10\%                &  \cite{Adam:2015vda}      \\
    $\bar{d}/d$              & $0.98 \pm 0.01 \pm 0.13$            & 0-10\%                &  \cite{Adam:2015vda}      \\
    $^3\overline{He}/^3He$   & $0.83 \pm 0.08 \pm 0.16$            & 0-20\%                &  \cite{Adam:2015vda}      \\
    $p$           &  $26 \pm 2.1$                                  & 0-20\%                &  \cite{Donigus:2015bsa}   \\
    $d$           &  $(8.71 \pm 0.04 \pm 1.58)\cdot 10^{-2}$       & 0-20\%                &  \cite{Donigus:2015bsa}   \\
    $^3He$             &  $(2.76 \pm 0.09 \pm 0.62)\cdot 10^{-4}$  & 0-20\%                &  \cite{Donigus:2015bsa}   \\
    $^4\overline{He}$  &  $(7.88 \pm 3.03 \pm 2.68)\cdot 10^{-7}$  & 0-20\%                &  \cite{Donigus:2015bsa}   \\
    $^{3}_{\Lambda}H$              & $(3.86 \pm 0.77 \pm 0.68) \times 10^{-5} \times$ B.R. & 0-10\%     &  \cite{Adam:2015yta}      \\
    $^{3}_{\bar{\Lambda}} \bar{H}$ & $(3.47 \pm 0.81 \pm 0.69) \times 10^{-5} \times$ B.R. & 0-10\%     &  \cite{Adam:2015yta}
  \end{tabular}
  \caption{Collection of $\frac{dN}{dy}$ of hadrons at $\sqrt{s_{NN}} = 2.76$ TeV  in Pb + Pb collisions measured by ALICE. 
  If two errors are given, then the first one is statistical, and the second one is systematic. B.R.\ denotes branching ratio of 
  $^{3}_{\Lambda}H \to ^3He + \pi^-$, which is estimated to be 15-35\%.}
  \label{tab:exp_data}
  \end{center}
\end{table}

\begin{table}
\begin{center}
  \begin{tabular}{ccc}
          Ratio                                              &   Value         & Error     \\
          \hline \\
          $\pi^-/\pi^+$                                      &        0.99776  &   0.10023 \\
          $K^-/K^+$                                          &        0.99500  &   0.11645 \\
          $\bar{p}/p$                                        &        0.98387  &   0.11313 \\
          $\Xi^-/\Xi^+$                                      &        1.01829  &   0.10552 \\
          $\Omega^-/\Omega^+$                                &        0.96667  &   0.23371 \\
          $\bar{d}/d$                                        &        0.98000  &   0.13038 \\
          $^3\overline{He}/^3He$                             &        0.83000  &   0.17889 \\
          $^{3}_{\bar{\Lambda}} \bar{H}/^{3}_{\Lambda}H$     &        0.89896  &   0.36364 \\
          $\phi/K^-$                                         &        0.11250  &   0.02500 \\
          $p/\pi^+$                                          &        0.04630  &   0.00500 \\
          $K^+/\pi^+$                                        &        0.14937  &   0.01605 \\
          $\Lambda/\pi^+$                                    &        0.03585  &   0.00453 \\
          $\Xi^+/\pi^+$                                      &        0.00490  &   0.00050 \\
          $\Omega^+/\pi^+$                                   &        0.00090  &   0.00016 \\
          $d/p$                                              &        0.00335  &   0.00067 \\
          $^3He/d$                                           &        0.00317  &   0.00092 \\
          $^4\overline{He}/^3He$                             &        0.00286  &   0.00207 \\
          $^{3}_{\Lambda}H/d$                                &        0.00177  &   0.00086
  \end{tabular}
  \caption{Ratios which are  analyzed here.}
  \label{tab:rat_fit}
  \end{center}
\end{table}

As usual, in all our fits of the ALICE data the finite width of resonances 
is always taken into account according to Eq.\ (\ref{EqIV}), while the  $\gamma_{S}$ parameter \cite{Rafelski}  
is set  $\gamma_{S}=1$ and all chemical potentials are set to zero.   As usual, the total multiplicities are found using 
the thermal and the decay contributions 
$n^{tot}_X = n^{th}_X+ n^{decay} = n^{th}_X + \sum_{Y} n^{th}_Y \, Br(Y \to X)$, where $Br(Y \to X)$ is the decay 
branching  ratio of  the Y-th hadron  into the hadron X (for more details see Ref.\ \cite{Veta14}). 
The expressions which are necessary to calculate the particle density of hadrons of species 
$k$ and their charge densities are given in the Appendix.

First we fitted the full set of ALICE data using the MHRGM for  the hard-core radii found earlier  \cite{Veta14}.   
From Fig.\ \ref{Fig3} one can see that the MHRGM describes the data a bit better than the ideal gas (where all hard-core 
radii are set to zero).
It is necessary  to note that  we did not include  five hard-core radii into  the number of degrees of freedom
since they were not used in the fitting procedure.
As one can see from  Fig.\ \ref{Fig3} the non-zero hard-core radii  help us to essentially improve  the description of
the $\frac{K^+}{\pi^+}$ ratio and slightly improve the $\frac{\Lambda}{\pi^+}$ ratio.

Similarly to Refs.\ \cite{Stachel:2013zma,Andronic16}  the MHRGM  is able to fit very well  the ratios involving (anti)nuclei  
(see Fig.\  \ref{Fig3}), however,
we believe that taking the hard-core radius of  (anti)nuclei to be the same as for baryons is not quite correct. 
The fact that the ideal gas is 
able to reproduce the ratios involving  (anti)nuclei  also requires an explanation. Therefore, in the rest of this work 
we compare the results which  include and the ones which exclude the (anti)nuclei 
ratios  into fits of the ALICE data. 
The latter option  is also  necessary  to qualitatively  compare our results with the model 
\cite{Vovch15} which does not include the (anti)nuclei data into a  fit. 
A typical example  of the IST EOS fit  results  without  the (anti)nuclei 
ratios    is  shown in Fig.\  \ref{Fig4}. It provides  the quality of the fit $\chi^2/ndf \simeq 8.04/10 \simeq 0.8$.

As one can see from Fig.\  \ref{Fig4}, due to an exclusion of (anti)nuclei data from the IST EOS  fit, the ratios 
$\frac{p}{\pi^+}$ and $\frac{\Xi^+}{\pi^+}$ are described slightly better  compared to the MHRGM, while the 
relative  deviation of the ratio $\frac{\Omega^+}{\pi^+}$  decreased from  the MHRGM value 1 to the IST EOS 
value 0.5.  Although the  ratio 
$\frac{K^+}{\pi^+}$ is better reproduced  by the MHRGM,
the quality of description of  all other ratios shown in 
Figs.\ \ref{Fig3} and \ref{Fig4} is almost the same. 
\begin{figure}
\centerline{
\includegraphics[width=\textwidth]{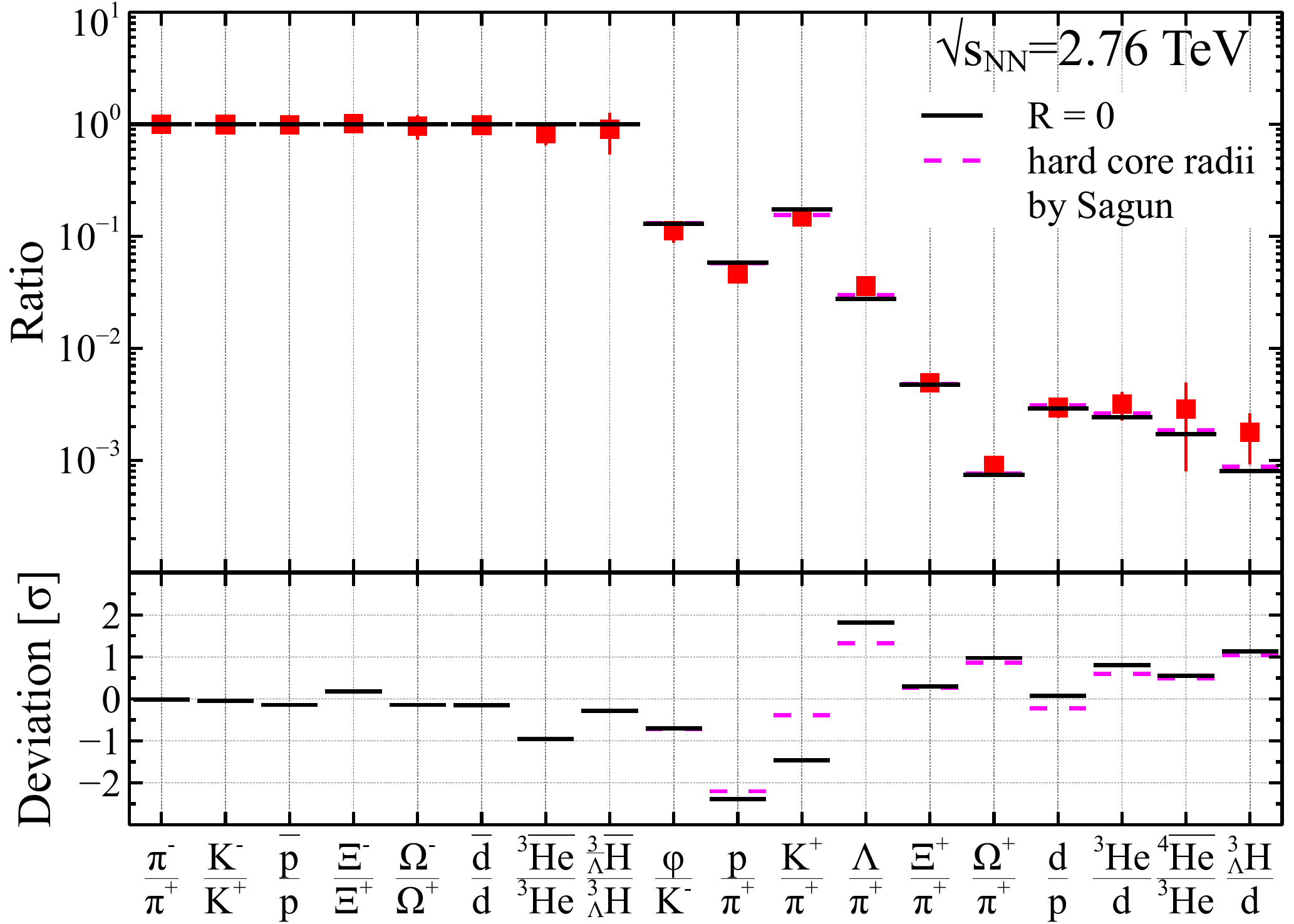}
}
 \caption{The full  set of  ALICE data (see Table  \ref{tab:rat_fit}) was fitted by the MHRGM with the hard-core radii  taken 
 from Ref.\ \cite{Veta14} with the CFO temperature $T_{CFO} \simeq 153 \pm 7$ MeV and  
 $\chi^2/ndf \simeq 10.9/17 \simeq 0.64$. For a comparison the ideal gas fit results are also shown which 
 correspond to $T_{CFO} \simeq 152 \pm 7$ MeV and $\chi^2/ndf \simeq 14.8/17 \simeq 0.87$.  
 The upper panel shows the fit of the ratios, while the lower panel shows the deviation between data and 
 theory in units of  estimated error.}
  \label{Fig3}
\end{figure}

\begin{figure}
\centerline{
\includegraphics[width=\textwidth]{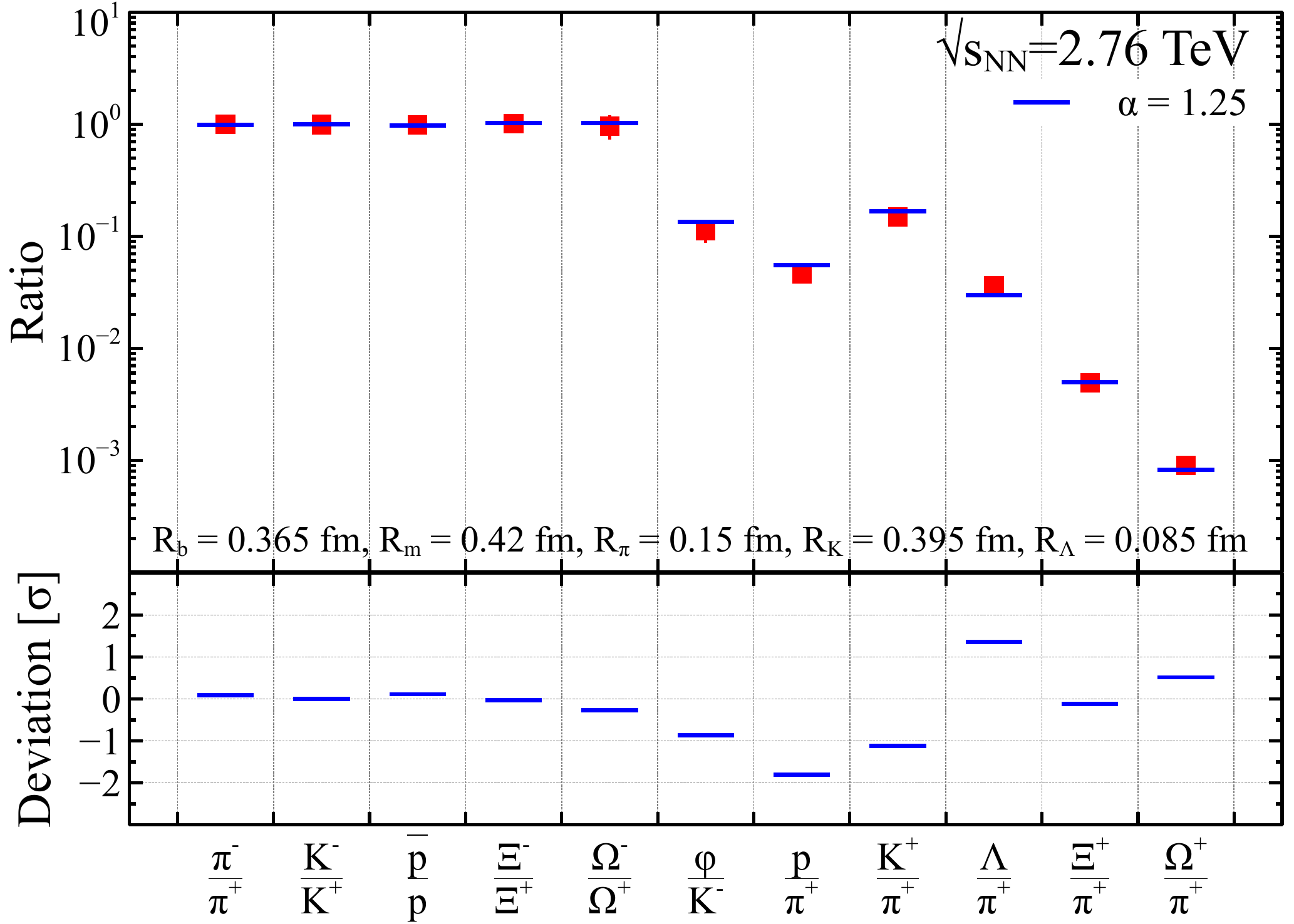}
}
 \caption{Same as in Fig.\ \ref{Fig3}, but the  fit was obtained by the IST EOS  with the new hard-core radii  found in 
 Ref.\ \cite{VetaNEW}.  The obtained  CFO temperature is  $T_{CFO} \simeq 152 \pm 7$ MeV. 
 The (anti)nuclei ratios are not included in the fit and its quality is 
 $\chi^2/ndf \simeq 8.04/10\simeq 0.8$. The upper panel shows the fit of the ratios, while the lower panel shows the 
 deviation between data and theory in units of  estimated error.}
\label{Fig4}
\end{figure}


\section{Analysis of  applicability range of various HRGM}

{
In the work  \cite{Vovch15} the fit of the ALICE data 
led to finding out of a second minimum of $\chi^2/ndf$ at the CFO temperature of 274 MeV. 
Therefore, here we would like to study the behavior of  $\chi^2 (T_{CFO})/ndf$
up to $T_{CFO} = 600$ MeV, but simultaneously we would like to determine the 
applicability range of each version of HRGM discussed above. 
Such a study is necessary because it is hard to believe that at so high densities provided by the  temperature  $274$ MeV  the chemical freeze-out can occur. 

The results of the fit with and without the light (anti)nuclei ratios are shown in Fig.\ \ref{Fig5} and  are summarized in Table 3. From Fig.\ \ref{Fig5}  one can see that the MHRGM and the IST EOS have a single minimum of $\chi^2 (T)/ndf$ for $T\le 600$ MeV.  These findings are in line with the results of the recent ALICE data analysis \cite{Andronic16}, although our values of $\min\{\chi^2 (T)/ndf \}$ are somewhat smaller.  Moreover,  as one can see from Table 3 the CFO temperatures of  all found minima agree with each other very well.  This means that the value of  CFO temperature is defined by the hadronic ratios. 

}

\begin{figure}[htbp]
\centerline{~~~
\includegraphics[width=84mm,height=77mm]{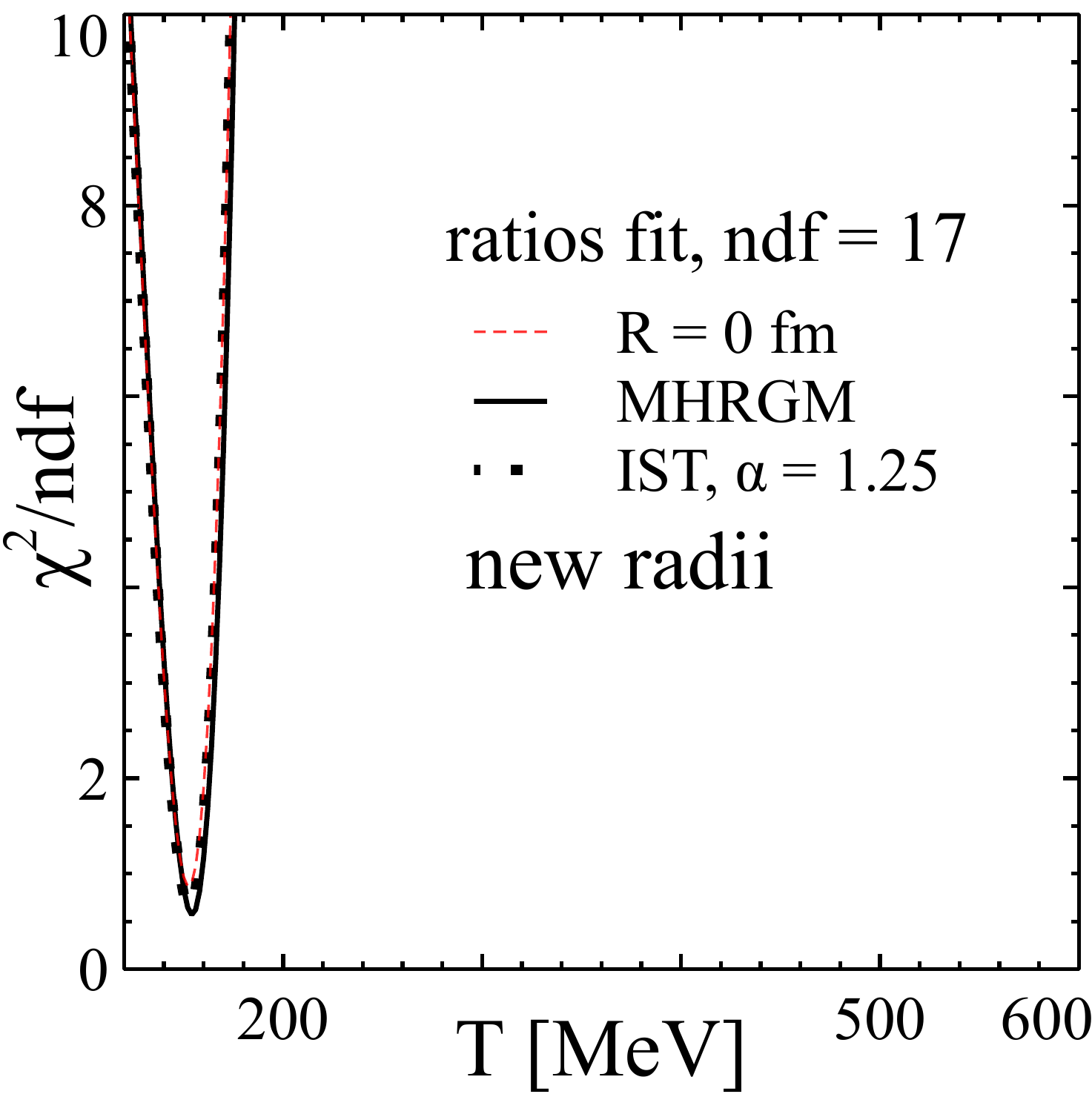}
  \hspace*{4.4mm}
\includegraphics[width=84mm,height=77mm]{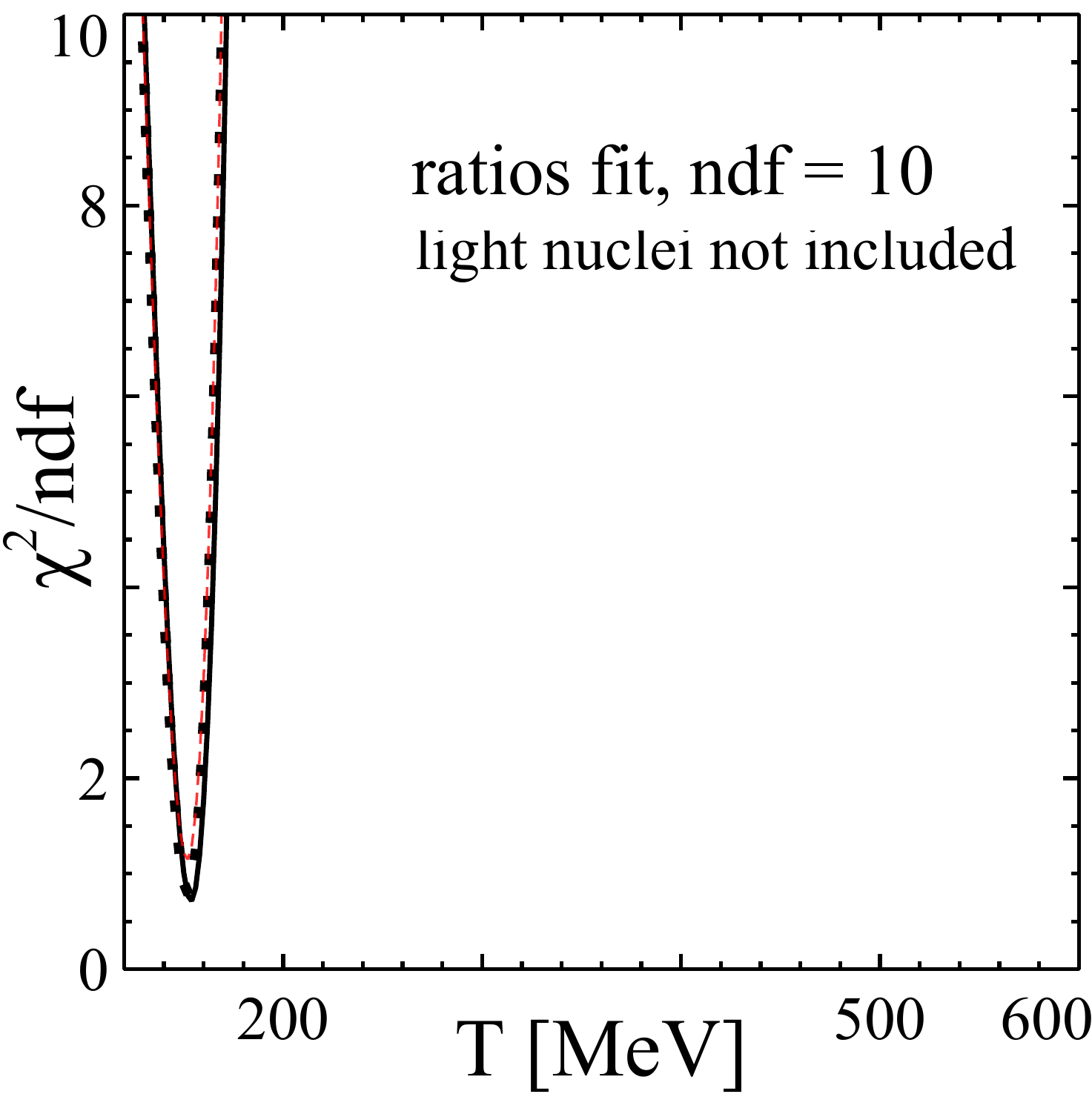}
}
 \caption{Comparison of the fit quality as a function of  CFO temperature  obtained for   the new hard-core radii  of  Ref.\ \cite{VetaNEW}  with  the  MHRGM (solid curve)  and with  IST EOS   (dotted  curve). 
The values of new hard-core radii  are given in Fig.\ \ref{Fig4}.
Also the results of the  fit  with the vanishing hard-core radii of all hadrons are shown by the dashed  curve. 
The left panel shows the results of fit in which   the hadron yield ratios and   the ones of light (anti)nuclei are taken into  account, while in  the right panel  only the hadron ratios are fitted.}
\label{Fig5}
\end{figure}

\begin{table}[ht]
\begin{center}
 \begin{tabular}{||c|c|c|c||}
 \hline
 \hline
        Ndf                 &   Model    &  $T_{CFO}$  (MeV)   & $\min\{\chi^2/ndf\}$   \\ 
                               &     & at $\min\{\chi^2/ndf\}$ & \\    \hline 
                          &  Ideal gas  &   $152\pm 7$ & $\frac{14.78}{17} \simeq 0.87$ \\
     17                     &  MHRGM  &   $154\pm 7$ & $\frac{9.71}{17} \simeq 0.57$ \\
                           & IST EOS &   $152\pm 7$ & $\frac{11.32}{17} \simeq 0.67$ \\                         
                       \hline    
                      \hline 
                          &  Ideal gas  &   $152\pm 7$ & $\frac{11.49}{10} \simeq 1.15$ \\
     10                     &  MHRGM  &   $154\pm 7$ & $\frac{7.16}{10} \simeq 0.72$ \\
                           & IST EOS &   $152\pm 7$ & $\frac{8.04}{10} \simeq 0.8$ \\                         
                       \hline    
                      \hline 

\end{tabular}     
\end{center}     
\vspace*{0.3cm}
\caption{
Parameters of $\chi^2/ndf$  minima for different versions of the HRGM found  for the new set of hard-core radii. 
The first column specifies the number of ratios used in the fit, the second column defines the  HRGM, 
the third column gives the CFO temperature at the $\chi^2/ndf$ minimum, whereas the last column gives
the value of   $\chi^2/ndf$ at the minimum. 
}
\label{table3}
\end{table}

Next we study the applicability of  various versions of  the  HRGM  at high temperatures. For this purpose we employ the multi-component version 
of the Carnahan-Starling EOS known as  the Mansoori-Carnahan-Starling-Leland (MCSL) EOS \cite{CSmultic}. 
This EOS is  well known  in the theory of simple 
liquids \cite{SimpleLiquids1} and has many applications in the statistical mechanics of polydisperse systems (see, for 
instance, Refs.\ \cite{SimpleLiquids2,LiquidsRev} and references therein). 
Similarly to the one-component case \cite{CSeos} the MCSL EOS accurately reproduces the pressure of hard spheres 
until packing-fraction values  $\eta \le 0.35-0.4$ \cite{CSmultic,SimpleLiquids2}. Here  the packing fraction of the 
$N$-component mixture is defined in a standard way $\eta \equiv  \sum\limits_{k=1}^N \frac{4}{3} \pi R_k^3 \rho_k$
via the set of  hard-core radii $\{ R_k\}$ and the corresponding particle densities $\{ \rho_k\}$.  In terms of these 
notations  the MCSL  pressure   \cite{CSmultic} reads 
 \begin{eqnarray}\label{EqXII}
p^{CS} &=&  \frac{6\, T}{\pi} \left[  \frac{\xi_0}{1-\xi_3}  + \frac{3\, \xi_1 \xi_2}{(1-\xi_3)^2} + 
\frac{3\,  \xi_2^3}{(1-\xi_3)^3}  -  \frac{ \xi_3  \xi_2^3}{(1-\xi_3)^3} \right] \,,\\
\label{EqXIII}
\xi_n  &=&   \frac{\pi}{6}  \sum\limits_{k=1}^N \rho_k \left[ 2\, R_k \right]^n \,.
\end{eqnarray}
{Using Eqs.\ (\ref{EqXII}) and (\ref{EqXIII}) we can determine the applicability bounds of  any version of the HRGM 
by comparing its  pressure with the MCSL pressure (\ref{EqXII}) found for the  same  value of temperature and for the same 
set of  particle densities  $\{\rho_k\}$. 
The results for the compressibility $Z = p/(\rho\, T)$ of the MHRGM   are depicted in 
Fig.\ \ref{Fig6}. Here 
the total  pressure of the system is $p$, while the total particle density is $\rho =  \sum\limits_{k=1}^N \rho_k$.  

In the left panel of  Fig.\ \ref{Fig6} we present a comparison for the new set of radii, while its right panel shows the results for the 
Vovchenko-Stoecker (VS)  \cite{Vovch15}  or the Bag model  prescription    
for the hard-core radii 
\begin{eqnarray}\label{EqXI}
R_k = R_0  \left[ \frac{m_k}{m_0} \right]^\frac{1}{3} \,, 
\end{eqnarray}
where   the constants are, respectively,   the hard-core radius  $R_0 = 0.5$ fm and the mass $m_0 =  938$ MeV of 
the proton \cite{Vovch15}. In contrast to all previous findings such a parameterization  led the authors of  
Ref.\  \cite{Vovch15} to the conclusion about existence of  a deeper minimum of  $\chi^2/ndf$
at very high temperature $T \simeq 274$ MeV where, according to present knowledge, 
the hadron gas does no longer exist.
Leaving aside the questions whether the prescription (\ref{EqXI}) can, in principle,  be applied to hadrons and how 
the CFO can occur  at  such a high temperature, we would like to  determine whether the  $\chi^2/ndf$ minimum at 
high temperatures is an artifact of  the EVM or it, indeed, has some physical meaning.  
}

\begin{figure}[htbp]
\centerline{~~~
\includegraphics[width=84mm,height=75mm]{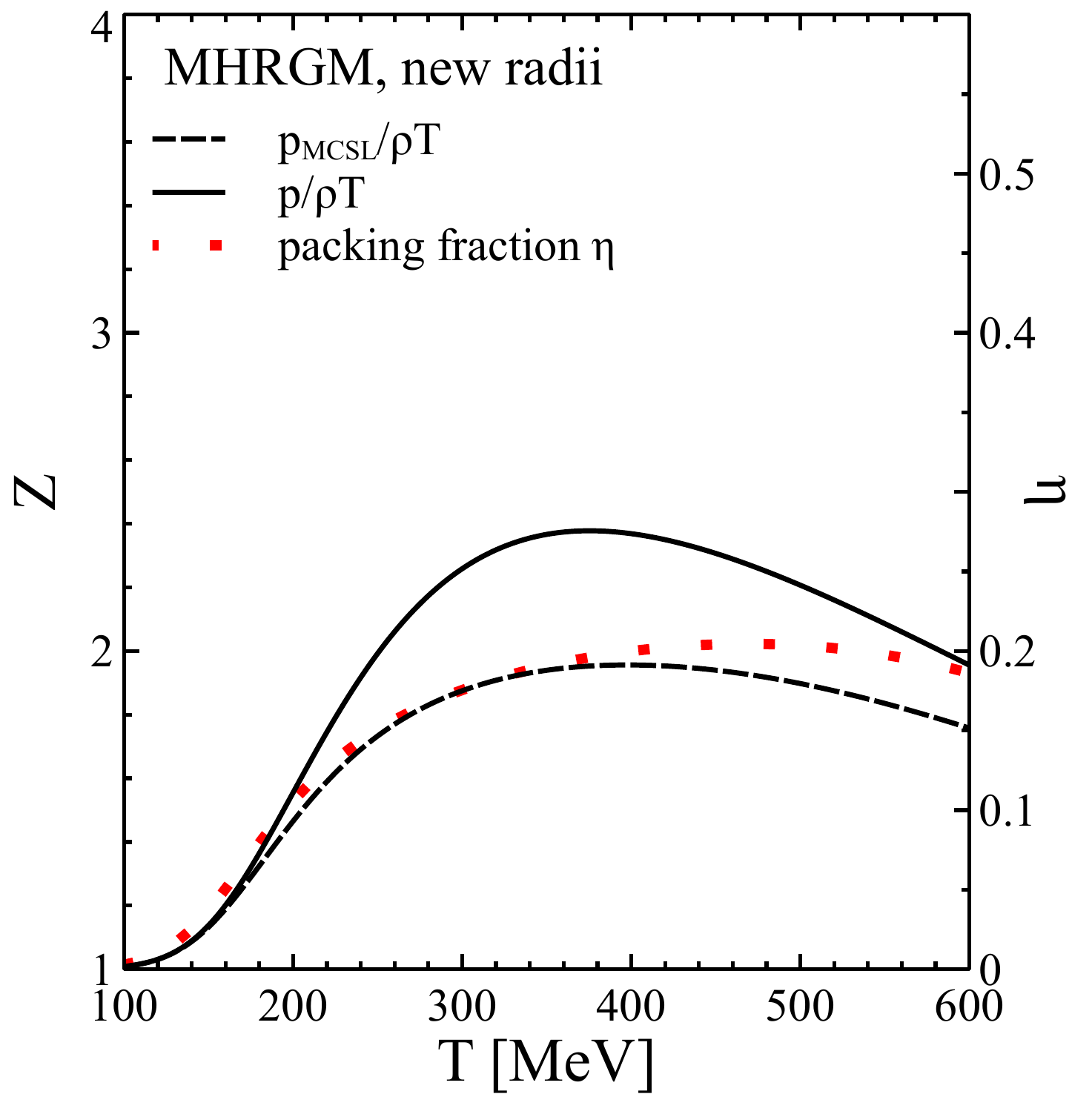}
  \hspace*{4.4mm}
\includegraphics[width=84mm,height=75mm]{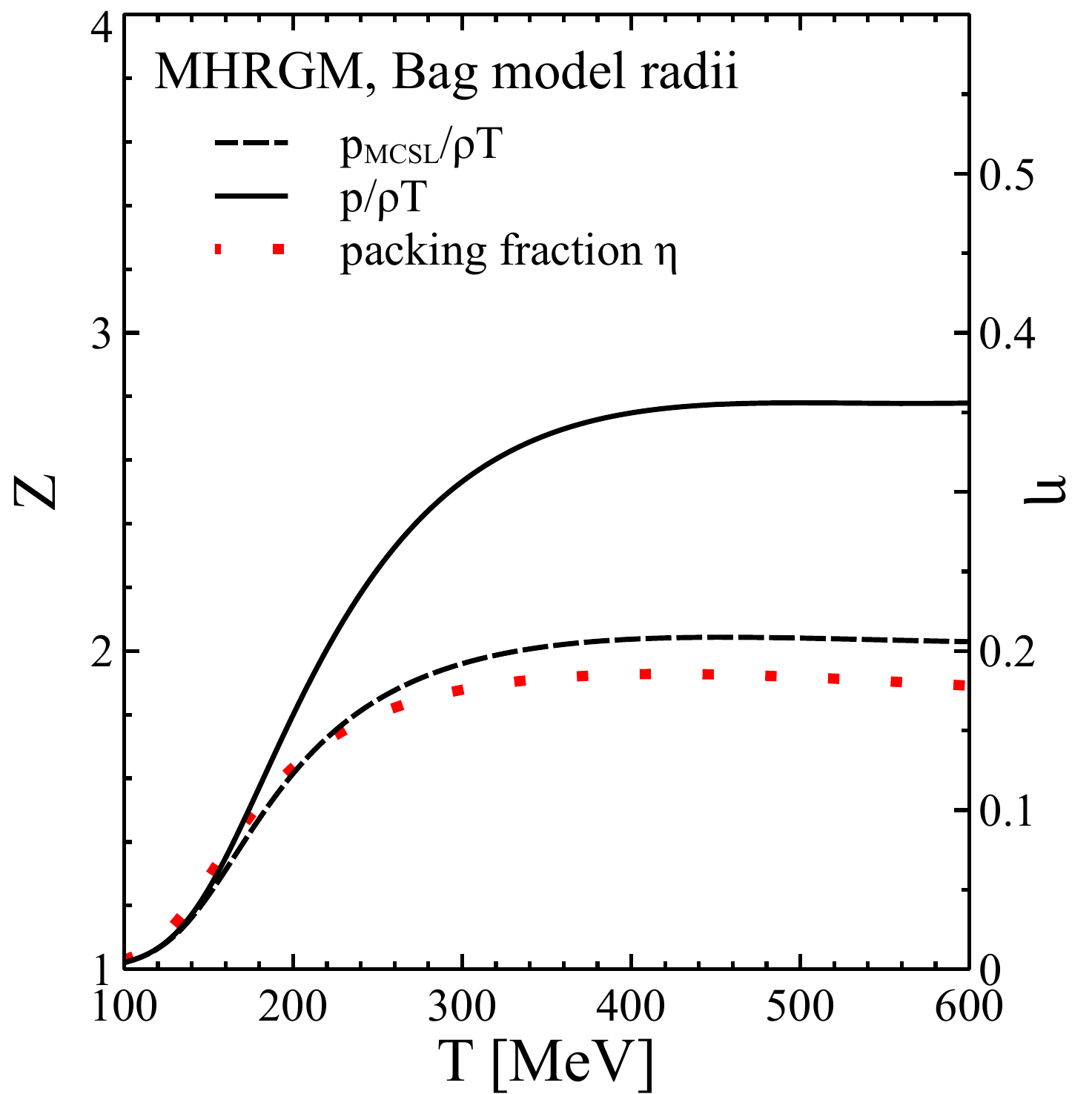}
}
 \caption{Thermal compressibility $Z$ as a function of  CFO temperature $T$  is shown for 
 the  MHRGM  and for    the MCSL EOS.  $Z$ are found  for the same particle densities. 
 {\bf Left panel:} the MHRGM (solid curve) and  the MCSL EOS (dashed curve) results   are 
 obtained for the new set of hard-core radii   \cite{VetaNEW}. 
 The dotted curve demonstrates the   CFO temperature dependence of the packing fraction.
 {\bf Right panel:}  
same as in the left panel, but  the VS  prescription of Eq.\ (\ref{EqXI})
(solid curve)  is compared with  the MCSL EOS (dashed curve) for the same particle densities. 
}
\label{Fig6}
\end{figure}
\begin{figure}[htbp]
\centerline{~~~
\includegraphics[width=84mm,height=75mm]{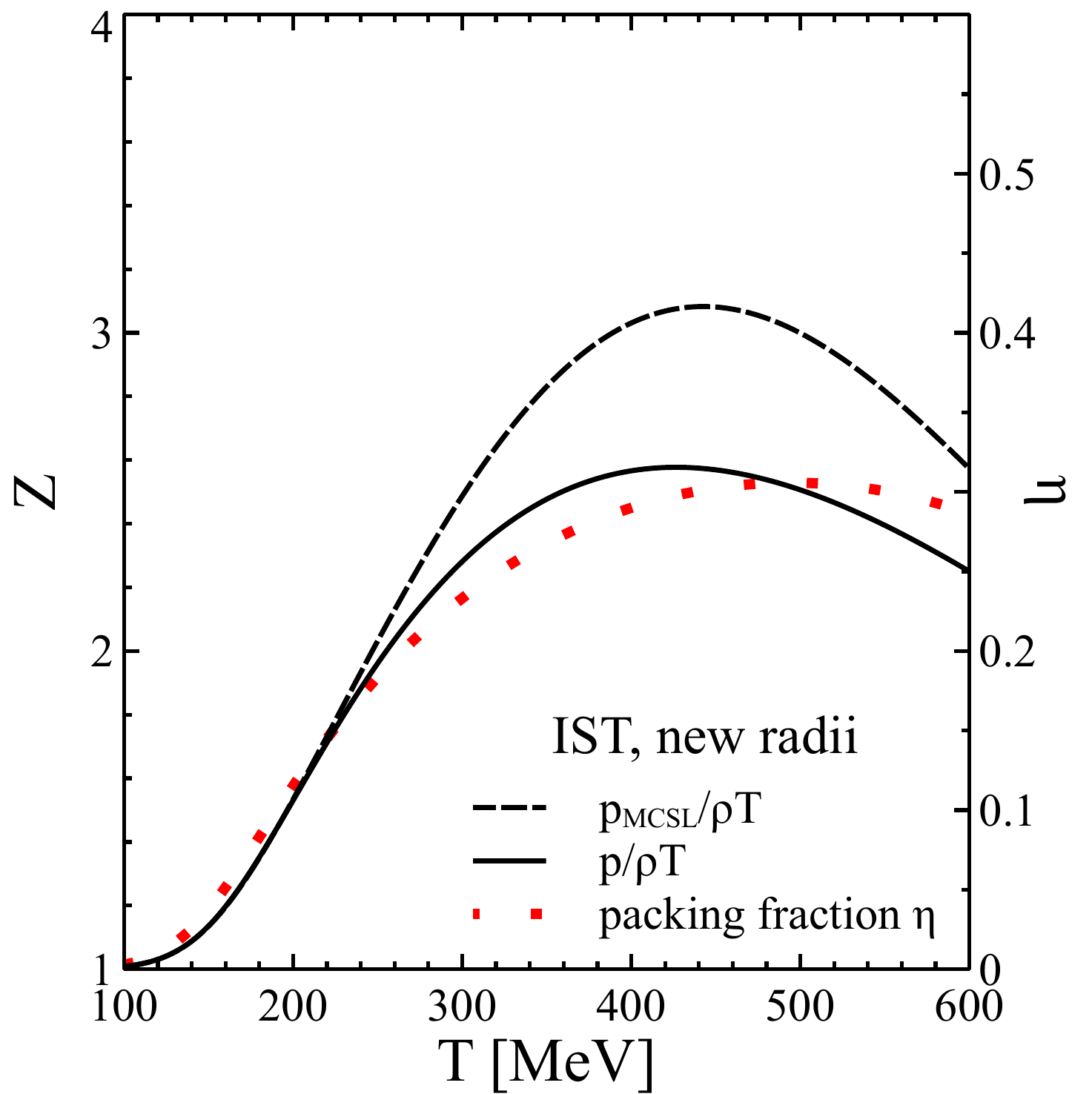}
  \hspace*{4.4mm}
\includegraphics[width=84mm,height=75mm]{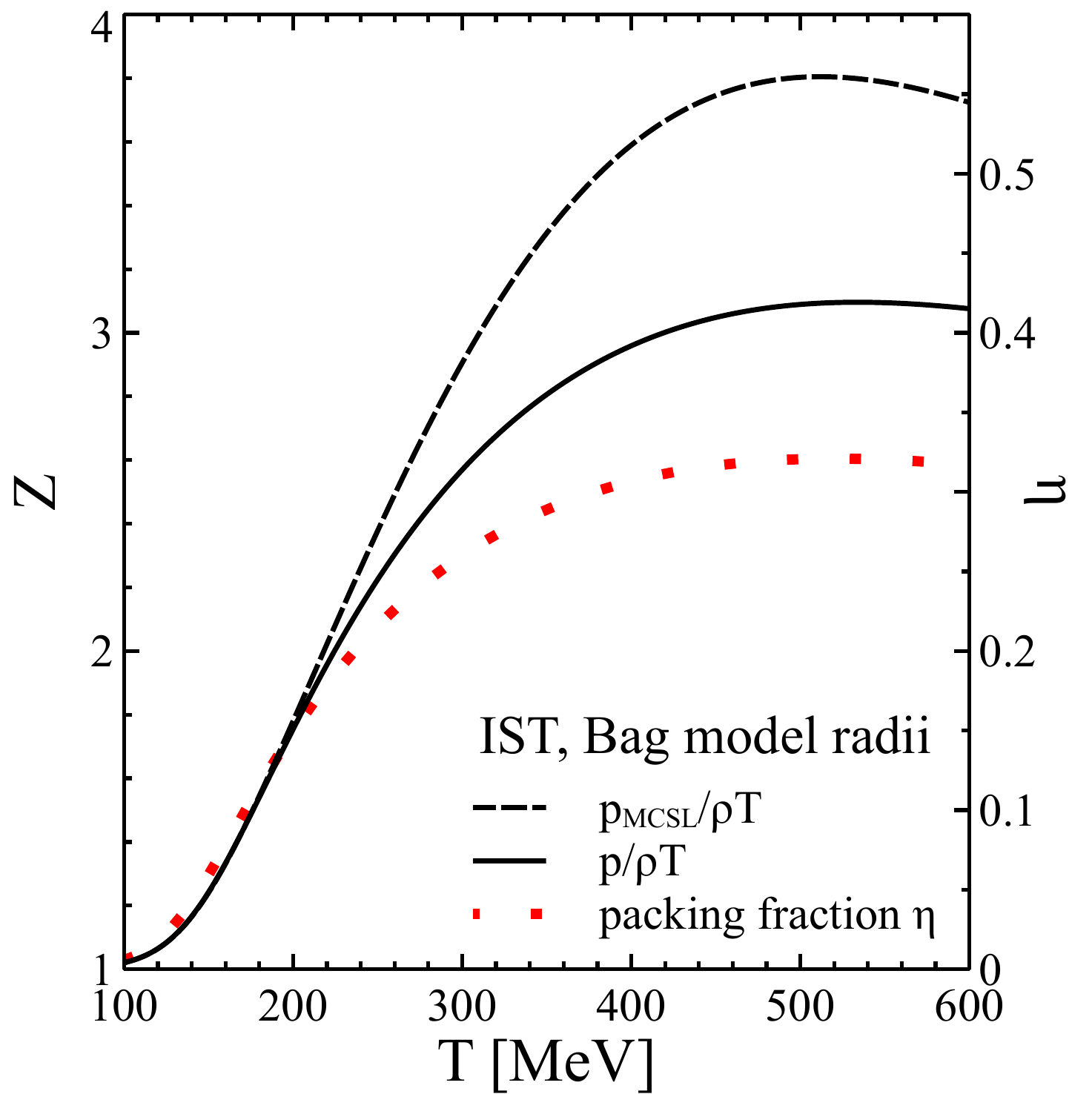} } 
 \caption{{\bf Left panel:} Same as  in Fig.\ \ref{Fig6}, but for the IST EOS with the new hard-core radii found in Ref.\
 \cite{VetaNEW}.
 {\bf Right panel:}  Same as  in Fig.\ \ref{Fig6}, but for the IST EOS with the VS prescription (\ref{EqXI}) for 
 the hard-core radii.}
  \label{Fig7}
\end{figure}

{Based on  the above   discussion of the EVM applicability (see Sec.\ 2),  we  defined it  by the inequality $\eta \le 0.11$ for  the packing fraction $\eta$.  This condition defines the applicability temperature $T_{appl} \simeq 200$ MeV for the MHRGM with the new radii (see the left panel of  Fig.\ \ref{Fig6}) and  $T_{appl} \simeq 180$ MeV for the VS model (see the right panel of  Fig.\ \ref{Fig6}).  Note that such an  inequality corresponds to 
about 6\% of  relative deviation between the compressibility factor $Z$ of the MHRGM and the one of the MCSL EOS. 
In other words,  the inequality $\eta \le 0.11$  is not too strict constraint  on the applicability bounds of the MHRGM. 

Also we  would like to point out that for $\eta \ge 0.175$ the both  versions of MHRGM  whose packing fractions are  shown in  Fig.\ \ref{Fig6}  become acausal, i.e. their speed of sound exceeds the speed of light. This finding  is similar to the result of VS model reported in \cite{Vovch15}. Thus,  the causality condition provides  a less strict  constraint on  the applicability bounds.  

A similar investigation  we performed for the IST EOS with the new set of hard-core radii and with the VS hard-core radii (\ref{EqXI}).
The results are shown in Fig.\  \ref{Fig7}.  Since the IST EOS is valid for the packing fractions obeying inequality $\eta \le 0.22$, we use this condition to determine the applicability temperature of this  EOS.  Using it we found that $T_{appl} \simeq 275$ MeV for the IST EOS with the new hard-core radii, while for the IST EOS with the VS  hard-core radii we got   $T_{appl} \simeq 235$ MeV. Note that  these values of the applicability temperatures provide about 5\% of  relative deviation between the compressibility factor $Z$ of the IST EOS  and the one of the MCSL EOS, i.e.  a constraint  $\eta \le 0.22$ is not too strict for
the IST EOS.  

 From this analysis  we can conclude  that  the minima of $\chi^2/ndf$ of the MHRGM  and the ones of the IST EOS  shown 
in Fig.\ \ref{Fig5} are located well inside the applicability bounds of these models.  
}

\section{Discussion of the results}

\begin{figure}[htbp]
\centerline{~~~
\includegraphics[width=84mm,height=75mm]{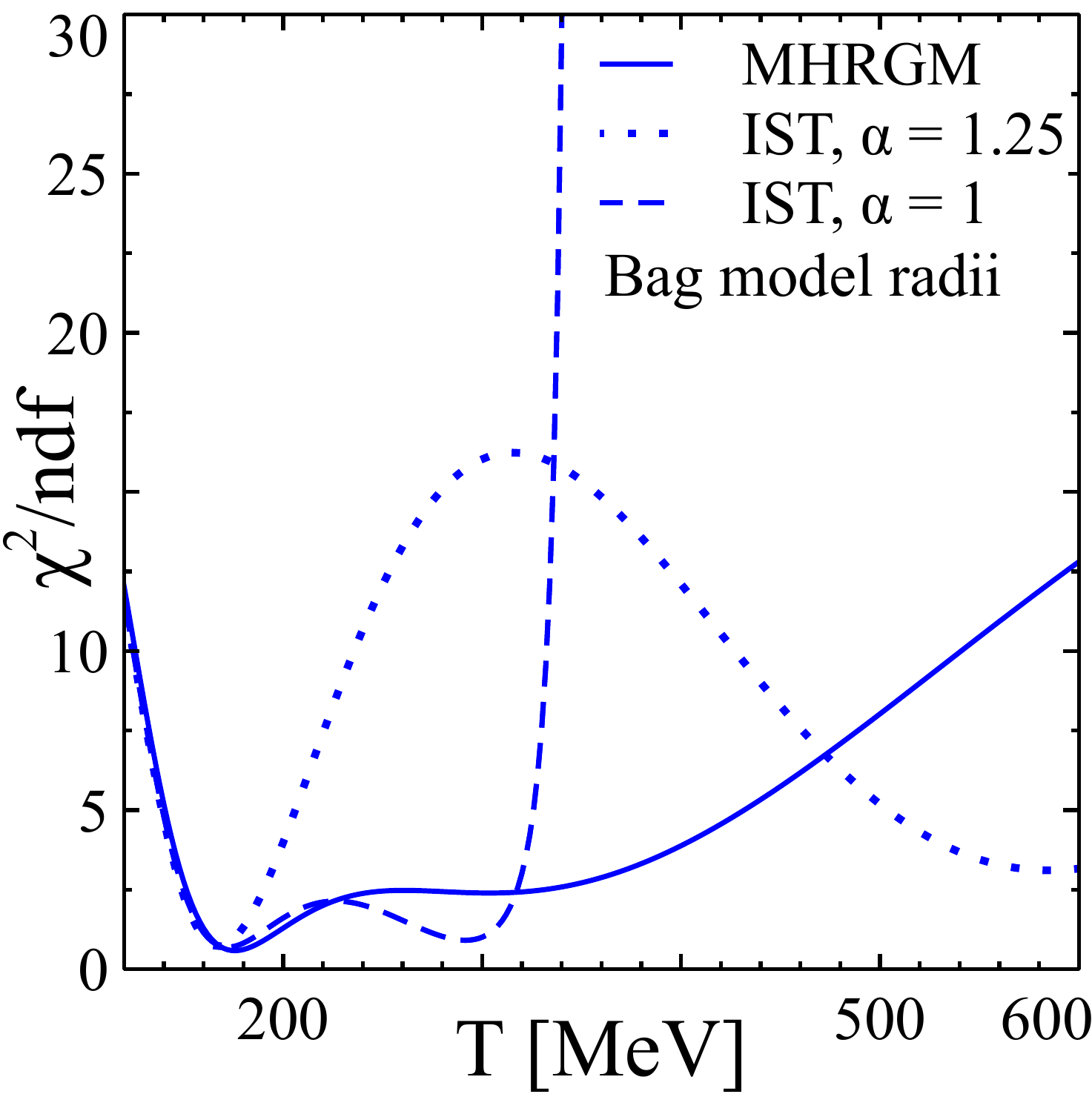}
  \hspace*{4.4mm}
\includegraphics[width=84mm,height=75mm]{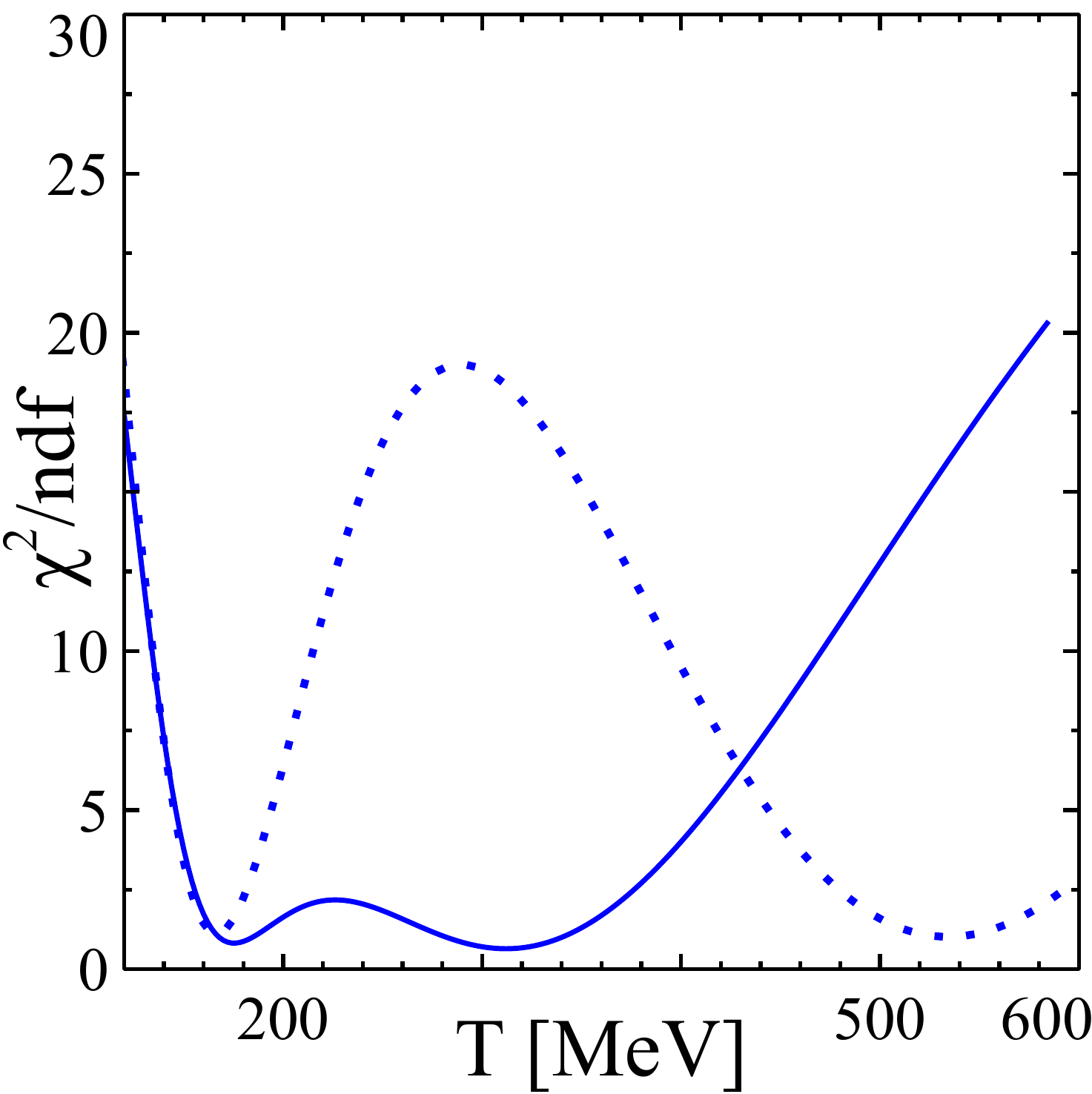}
}
 \caption{The fit quality as a function of  CFO temperature for 
 the MHRGM  (solid curve), IST EOS with $\alpha =1.25$ (dotted curve) and  IST EOS with $\alpha =1$ (dashed curve).
 All models  employ  the VS prescription of Eq.\ (\ref{EqXI}) for  hard-core radii.
Left panel shows the results for the light (anti)nuclei ratios included into the fit, while  the results shown in  right panel do not include  
these  ratios.  Note that for $T \le 230$ MeV the solid and long dashed curves are hardly distinguishable  because these are EVM.}
\label{Fig8}
\end{figure}
\begin{table}[htb]
\begin{center}
 \begin{tabular}{||c|c|c|c|c|c||}
 \hline
 \hline
        Ndf                 &   Model    &  $T^{low}_{CFO}$  (MeV)   & $\min\{\chi^2/ndf\}$   &  $T^{high}_{CFO}$  (MeV)   & $\min\{\chi^2/ndf\}$   \\ 
                               &     & at $\min\{\chi^2/ndf\}$ &  & at $\min\{\chi^2/ndf\}$ & \\    \hline 
&  MHRGM  &   $176^{+30}_{- 18}$ & $\frac{9.92}{17} \simeq 0.58$  & $304^{+78}_{-43}$   & $\frac{40.71}{17} \simeq 2.4$ \\ 
 17 & IST EOS  $(\alpha=1)$&   $172^{+30}_{-19}$ & $\frac{11.78}{17} \simeq 0.69$  &  $292^{+21}_{-46}$    & $\frac{15.48}{17} \simeq 0.91$ \\             
 & IST EOS &   $168\pm 14$ & $\frac{12.36}{17} \simeq 0.72$  & $586 \pm 60$   &   $\frac{52.82}{17} \simeq 3.1$\\            
                       \hline    
                      \hline 
  &  MHRGM  &   $176^{+30}_{-17}$  & $\frac{8.21}{10} \simeq 0.82$ & $312^{+46}_{-54}$ & $\frac{6.44}{10} \simeq 0.64$ \\
   10  &  & & & & \\
                           & IST EOS &   $166^{+14}_{-13}$  & $\frac{10.13}{10} \simeq 1.01$ &  $534^{+52}_{-48}$ & $\frac{9.18}{10} \simeq 0.92$ \\                         
                       \hline    
                      \hline 

\end{tabular}     
\end{center}     
\vspace*{0.3cm}
\caption{
Parameters of $\chi^2/ndf$  minima for different versions of the HRGM found for the VS hard-core radii. 
The first, second, third and fourth   columns have the same meaning as the ones in Table 3.
The fifth column gives the CFO temperature at the second minimum of  $\chi^2/ndf$, and  the last column gives
the value of   $\chi^2/ndf$ at this  minimum. 
}
\label{table4}
\end{table}

Now we are ready to apply the IST EOS to the description  of the ALICE data with the 
VS  prescription for hard-core radii. 
For such a study     we extended   the MHRGM  \cite{Horn,SFO,Veta14,Bugaev:2014} to 
the  VS prescription  (\ref{EqXI}) and fitted  the ratios of hadron multiplicities.  
In other words,  using  Eq.\ (\ref{EqXI})  we calculate the second virial coefficient between the particles $k$ and $l$
with the hard-core repulsion  as usual $b_{kl} = \frac{2\, \pi}{3} (R_k + R_l)^3$.

{The obtained  results are presented  in Fig.\ \ref{Fig8}.
This figure shows the CFO temperature dependence of the fit quality $\chi^2/ndf$. In this fit we used  
the same parameters in Eq.\ (\ref{EqXI}) as in Ref.\  \cite{Vovch15}.  From  Fig.\ \ref{Fig8} one
can see that the VS prescription for hard-core radii always generates the second minimum of  $\chi^2/ndf$
at high temperatures. The parameters of  all minima shown in   Fig.\ \ref{Fig8} 
are given in Table 4.  

We confirm the finding of Ref. \cite{Vovch15} that the minima of the VS prescription  (\ref{EqXI}) 
are strongly asymmetric for the MHRGM. The same is true for the IST EOS with $\alpha =1$, which, as we discussed in Sec.\ 2, is 
also the EVM.  This statement becomes apparent from the left panel of Fig.  \  \ref{Fig7}, if one compares the temperature dependence of  solid and dashed curves in this figure.  On the other hand,  the  IST EOS with $\alpha =1.25$ 
demonstrates  essentially  different $T$-behavior (see below)   and the asymmetry of its  minima is less pronounced. 
Also the CFO temperatures of the first  $\chi^2/ndf$ minimum (see the third column in Table 4) are shifted to higher temperatures  
166-176 MeV  compared to the case of the fixed  hard-core radii. This is also in line with  Ref. \cite{Vovch15} results for the MHRGM. 

As one can see from Table 4, if  the  light (anti)nuclei are included into a fit,  the minimum at higher temperature is  shallower, whereas, if they are not included,  the minimum at higher temperature is slightly deeper. This is similar to the results of Ref. \cite{Vovch15} for the MHRGM.  However, the main differences between the results of the EVM and the IST EOS are as:\\
(i)  the high $T$ minimum is shifted to temperatures above 500 MeV at which, according to the present days paradigm,  one cannot use the hadronic language at all  and the CFO concept does not make any sense, since the inelastic reactions at these temperatures are not frozen;\\
(ii) in contrast to the EVM results,  the two minima  of the IST EOS  $\chi^2/ndf$ are separated by a huge maximum.

Therefore, the expectations of the authors of Ref. \cite{Vovch15},  that the more elaborate models will confirm their statement that the 
CFO may occur in the  temperature  range 170-320 MeV, are not supported by our results. Moreover, comparing the applicability temperature  $T_{appl} \simeq 235$ MeV  of  the IST EOS which employed  the  VS prescription with the temperatures of the $\chi^2/ndf$ second   minimum (see the fifth column in Table 4), we conclude that none of the considered models can be applied at the temperatures at which the second minimum is found.  Hence, we do not think it makes any sense to owe any physical  meaning to  these high temperature minima.  Our second  major conclusion is that the second minimum found in Ref.\  \cite{Vovch15}  at $T\simeq 274$ MeV and observed  here  in the  EVM at  $T\simeq 292-312$ MeV  are artifacts of 
 extrapolating   the EVM  with  the VS prescription   (\ref{EqXI}) for hard-core radii 
 far beyond the limit of its applicability.  
}

An apparent reason of the EVM  failure 
at  temperatures above 200  MeV is that  it overestimates the third and higher virial coefficients 
compared to a more elaborate MCSL EOS and the IST EOS (the third and fourth virial coefficients of the one component
IST EOS can be found in \cite{Sagun16}). 
Furthermore, it is clear that at high temperatures 
one cannot ignore the Lorentz contraction of the hard spheres \cite{RelVDW1,RelVDW2,RelVDW3,RelVDW4}, since the 
existence of hard spheres is in contradiction with the postulates of  relativity.  Therefore, strictly speaking,  all results 
of the EVM which are 
shown in Fig.\ \ref{Fig6} cannot be considered as trustworthy at temperatures above 200 MeV \cite{RelVDW1}.  
The effect of Lorentz contraction leads to  a decreasing of the mean hard-core radius with the temperature,  
but for temperatures obeying the inequality $m_h < T < 2 m_h$ (here $m_h$ is the mass of hadron)
the mean hard-core radius  of a hadron is  $2^\frac{1}{3} \simeq 1.26$ times smaller than the one at vanishing 
temperature \cite{RelVDW1}.  Such a decrease  corresponds to an about 50\% reduction of the excluded volume of 
two identical non-relativistic hard spheres \cite{RelVDW1}.  In other words, within the EVM  for  $T \sim 250$ MeV 
one has to account for the reduction of the hard-core radii of pions, kaons, and $\eta$ meson, while for  $T \sim 350$ MeV 
one should in addition account for such a reduction of  the  $\omega$ and $\rho$ mesons' hard-core radii etc. 

A principally different situation occurs for the IST EOS with $\alpha =1.25$ and with the  fixed hard-core radii of hadrons, since it correctly reproduces the MSCL EOS at those temperatures where the EVM fails, as one can see from the left panel of  Fig.\ \ref{Fig7}. 
Employing  the new hard-core radii in  the IST EOS, one finds that this  EOS provides not more than  a 5\% deviation
from the MSCL EOS at $T \le  T_{appl} \simeq 275$ MeV, i.e., in the region where the second minimum of the VS model was found in \cite{Vovch15}. 
{From the discussion  given in Sec.\ 2  it is clear that the IST EOS accounts for the gradual decrease of the excluded volume of hadrons when the pressure  increases.  In principle,  one should not consider  the temperatures above $T_{appl}$, but let us
discuss what will happen at higher temperatures.
As one can see from the left panel of  Fig.\ \ref{Fig7}, at $T \simeq 340-350$ MeV
 the regime of the proper volume is reached by almost all  hadrons and, hence, the packing fraction saturates at the value $\eta \simeq 0.3$. 
 Accounting for this 
fact and remembering that the pion hard-core radius $R_\pi =0.15$ fm is almost three times smaller than the hard-core radii 
of   other mesons (hence the pion proper volume is about 25 times smaller than the proper volume of  other mesons),  
we conclude: (i) at such temperatures the pions behave almost as point-like particles; (ii) the excluded volumes of other hadrons are now reduced to their proper volumes which are small. For instance, the proper volume of 
all baryons $V_b^{new}$, except the $\Lambda$-hyperons,  is about $V_b^{new} \simeq 0.2$ fm$^3$.
Hence,  accounting for the relativistic effects  of a few species (kaons,  $\eta$, $\omega$ and $\rho$ mesons etc. )  will  stronger  approximate the  hadronic pressure to the one of  ideal gas, which does not  generate the second minimum at high temperatures. 
Thus, the Lorentz contraction effect cannot  strongly affect the results of the IST EOS with the fixed values of  hard-core radii 
 \cite{VetaNEW}.
}

{Using the VS prescription (\ref{EqXI}) within the IST EOS we found its applicability temperature $T_{appl}$ which  provides not more than a 5\% deviation from the MSCL EOS at 
$T_{appl} \simeq 235$ MeV. Note that the latter temperature  is far below  the region in which the authors of Ref.\ \cite{Vovch15}
found the second minimum of $\chi^2/ndf(T)$ and, hence, at this point  one could stop any discussion of this issue. 
Moreover,  according to the lattice QCD  one cannot use the hadronic language at temperatures above 160-170 MeV at any baryonic density and, therefore,  we do not have to discuss the VS prescription further.  However, let us see what will happen,  if one  follows the logic of Refs. \cite{Vovch15,Vovch16}   insisting  on the fact that  the IST EOS with 
the VS hard-core radii provides  a better description of the present ALICE data  at the CFO temperatures above 500 MeV. 
In this case  at $T \ge 500$ MeV only pions, $\eta$-mesons and kaons will behave as the ideal gas, whereas 
all heavier hadrons  will have rather large proper volumes.
For instance,    for    the proton radius $R_0 =0.5$ fm used in \cite{Vovch15} the nucleon  proper  volume  $V_p^{VS}$ in the VS model  is $V_p^{VS} \simeq 0.52$ fm$^3$ which is essentially larger than the    proper  volume  of  baryons   $V_b^{new}$   found for the set of new hard-core radii \cite{VetaNEW}.
Therefore,  for nucleons and  heavier hadrons  one will have  to include the Lorentz contraction of  the  VS hard-core radii. 

Note that  at $T=500$ MeV  the nucleon  proper  volume  $V_p^{VS}$  is reduced to about $0.65 \,V_p^{VS} \simeq 0.34$ fm$^3$ due to the  Lorentz contraction  \cite{RelVDW2}. Similarly, the effect of  induced surface tension  of nucleons and heavier hadrons will  weaken  and, hence,  the number of all these hadrons will be  sizably enhanced due to high pressure. 
Using the qualitative explanation of the fact that  decreasing the hard-core radius of nucleons, one  increases  the temperature of the second minimum of $\chi^2/ndf$ for the ALICE data   \cite{Vovch16,Satarov16}, 
we conclude that inclusion of  the Lorentz contraction into the 
IST EOS with VS radii  will inevitably shift  the second minimum to even higher temperatures 
which may not be achieved  by the initially thermalized  state  formed in relativistic heavy-ion collisions.
In other words,  in this section  it is  shown  that the location  of a 
minimum at high temperatures depends on the degree of model refinement.  
Then our  final conclusion is that the additional 
minimum, found in Ref.\ \cite{Vovch15}, is unphysical.
}

\section{Conclusions}

 In this work we developed 
a novel formulation of the HRGM, the so-called induced surface tension EOS (IST EOS), which
explicitly accounts for the  surface tension induced by the interaction between  particles. 
Such an approach was developed earlier in Ref.\ \cite{Bugaev:13NPA}  on the basis of the virial expansion for 
an $N$-component 
mixture of gases which have individual hard-core radii. Using the freedom of the Van der Waals extrapolation to high 
particle densities we introduced a parameter $\alpha$, which here  
is found to serve as a ``switch"  between  the excluded  and proper volume regimes. A  detailed comparison with the 
famous Carnahan-Starling EOS demonstrates the validity of the IST EOS  at higher packing fractions  0.2-0.22 than the 
traditional EVM, which is valid up to packing fractions of about 0.1-0.11. Moreover, we found that 
the IST EOS is softer than the Carnahan-Starling EOS \cite{CSeos} and its multi-component version 
MCSL EOS \cite{CSmultic};
as a result, the model  respects causality up to higher densities than these famous EOS. 

Here we  constructed the particle-yield ratios from the multiplicities measured by the ALICE experiment at 
$\sqrt{s_{NN}}=2.76$ TeV, which we  fitted to the different  versions of the HRGM, namely the EVM and  IST EOS. 
It is found that the MHRGM and  IST EOS provide almost the same high fit quality with $\chi^2/ndf \simeq 0.6-0.8$ and 
practically the same value of the  CFO temperature $T_{CFO } \simeq 154 \pm 7$ MeV. 
In addition, within the MHRGM  we analyzed the VS model with the  bag-like prescription (\ref{EqXI}) for the 
hard-core radii.  
Due to a different realization of the Van der Waals repulsion in our analysis of  the VS  prescription (\ref{EqXI}) we found 
a second minimum of $\chi^2/ndf$ at  slightly higher temperatures  $T = 292-312$ MeV compared to the work 
\cite{Vovch15}. 
Also  we showed that the MHRGM with  the VS  prescription (\ref{EqXI})  for hard-core radii cannot be applied at 
temperatures  above $T \simeq 180$ MeV, since it gets stiffer than the non-relativistic mixture of hard spheres.  
An apparent consequence is that its speed of sound exceeds the speed of light
at high temperatures, which makes the whole treatment 
unphysical.

Using the IST EOS we performed the fit of ALICE data with fixed values of hard-core radii for pions, kaons, $\Lambda$ 
(anti)hyperons, other mesons, and other baryons and found no traces of the high-temperature minimum for $T_{CFO} \le 600$ MeV, although such a 
model, as we showed here, is applicable at temperatures below 275 MeV.  Furthermore, to get rid of any suspicions about 
existence of the high-temperature minimum for the VS  prescription (\ref{EqXI}), we analyzed the ALICE data with
the IST EOS using hard-core radii given by Eq.\ (\ref{EqXI}). Although such an EOS is applicable at temperatures 
below 235 MeV only,  we did not see any  additional minimum within this range of 
 temperatures, whereas  we again found a second minimum at a temperature above 500 MeV, i.e., 
 where the IST EOS is inapplicable. Therefore, we showed that the high-temperature  minimum of $\chi^2/ndf$ found in 
 Ref.\ \cite{Vovch15} and further discussed in \cite{Vovch16}  is a consequence 
 of extrapolating the Van der Waals EOS far beyond the limits of its applicability.

\vspace*{2.2mm}

{\bf Acknowledgments.}
The authors  are   thankful to A. Andronic, D. B. Blaschke, P. Braun-Munzinger, T. Galatyuk,  C. Greiner,  and M. Gazdzicki  
for fruitful discussions and for important  comments. 
 We thank V. Vovchenko for the constructive criticism, comparison to his results and important corrections.
The authors give their cordial thanks to D. R. Oliinychenko for a through verification 
of fitting results, for very important critical comments  and for a suggestion to present  the packing fraction in the figures 6 and 7.  
We thank D. H. Rischke for a critical reading of the manuscript.
K.A.B., V.V.S., A.I.I. and G.M.Z. acknowledge a partial support from 
the program ``Nuclear matter under extreme conditions'' launched 
by the Section of Nuclear Physics of NAS of Ukraine. 
K.A.B. acknowledges a partial support by the 
ExtreMe Matter Institute EMMI, GSI Helmholtzzentrum f\"ur Schwerionenforschung, Darmstadt, Germany.


\vspace*{0.22cm}

\section{Appendix: Expression for particle density}

In order to compare the IST EOS with the generalized Carnahan-Starling EOS \cite{CSmultic}  we need to have the explicit 
expressions for the particle density of  hadrons   of species $k$. Here we consider the generalized system 
(\ref{EqI}) and (\ref{EqII}), i.e., the partial pressure  $p_k$ and the partial surface-tension coefficient $\Sigma_k$ are 
defined as
\begin{eqnarray}
\label{EqIA}
p_k &=& T \phi_k \exp \left[ \frac{\mu_k}{T} - \frac{4}{3}\pi R_k^3 \frac{p}{T} - 4\pi R_k^2 \frac{\Sigma}{T} \right]
\,, \\
\label{EqIIA}
\Sigma_k &=& T  R_k \phi_k \exp \left[ \frac{\mu_k}{T} - \frac{4}{3}\pi R_k^3 \frac{p}{T} - 4\pi R_k^2 \alpha_k 
\frac{\Sigma}{T} \right]  \equiv  p_k R_k  \exp\left[ - 4\pi R_k^2  (\alpha_k-1) \frac{\Sigma}{T}   \right] \,,
\end{eqnarray}
where the total chemical potential is given by Eq.\ (\ref{EqIII}). Then the total pressure and the total surface-tension 
coefficient are defined as $p = \sum_k p_k$ and  $\Sigma = \sum_k \Sigma_k$, respectively.
The  system   (\ref{EqIA}) and  (\ref{EqIIA}) is a generalization of Eqs.\  (\ref{EqI}) and  (\ref{EqII}) to the case when each 
particle species has its own value of the parameter $\alpha_k$. Evidently, setting 
$\alpha_1 = \alpha_2 = ... = \alpha_k = ...= \alpha$
in Eqs.\ (\ref{EqIA}) and  (\ref{EqIIA}) one obtains Eqs.\ (\ref{EqI}) and  (\ref{EqII}).

Differentiating $p$  and $\Sigma$   with respect to the full chemical potential $\mu_k$ of  the hadron of sort $k$  one finds 
\begin{eqnarray}
\left( \begin{array}{cc}
a_{11} & a_{12} \\
a_{21} & a_{22}
\end{array} \right) \cdot
\left( \begin{array}{c}
\frac{\partial p}{\partial \mu_k}  \\
\frac{\partial \Sigma}{\partial \mu_k}
\end{array} \right) = 
\left( \begin{array}{c}
\frac{p_k}{T} \\
\frac{\Sigma_k}{T}
\end{array} \right)
\end{eqnarray}
Here the coefficients $a_{kl}$ can be expressed in terms of  the partial pressures $\{ p_k\}$ and the 
partial surface-tension coefficients $\{\Sigma_k\}$ as
\begin{eqnarray}
a_{11} &=& 1 + \frac{4}{3} \pi \sum_k  R_k^3 \frac{p_k}{T} \,, \\
a_{12} &=& 4 \pi \sum_k R_k^2 \frac{p_k}{T} \, , \\
a_{21} &=& \frac{4}{3} \pi \sum_k R_k^3\frac{\Sigma_k}{T} \, ,\\
a_{22} &=& 1 + 4 \pi \sum_k  R_k^2 \alpha_k\frac{\Sigma_k}{T} \,.
\end{eqnarray}
Then the particle density of hadrons of species $k$ is  given by
\begin{equation}\label{EqVIIIA}
\rho_k \equiv \frac{\partial  p}{\partial \mu_k} = \frac{1}{T} \cdot \frac{p_k \, a_{22} 
- \Sigma_k \, a_{12}}{a_{11}\, a_{22} - a_{12}\, a_{21} } \,.
\end{equation}
The charge density of kind $A$ $(A \in \{B, S, I_3\})$ of a hadron of species $k$ can be found 
by multiplying Eq.\ (\ref{EqVIIIA}) by  the partial derivative $\frac{\partial \mu_k}{\partial \mu_A} = A_k$.


\end{document}